\documentclass[twocolumn]{aastex63}
\usepackage{color,soul}
\received{}
\revised{}
\accepted{}
\submitjournal{ApJ}

\shorttitle{Environmental Dependence of $z<$0.3 Quasars}
\shortauthors{Wethers et al.}
\graphicspath{{./}{}}

\begin{document}

\tolerance=1
\emergencystretch=\maxdimen
\hyphenpenalty=10000
\hbadness=10000

\title[Environmental Dependence of 0.1$<z<$0.35 Quasars]{Galaxy and Mass Assembly (GAMA): The Weak Environmental Dependence of Quasar Activity at 0.1$<z<$0.35}

\correspondingauthor{Clare F. Wethers}
\email{clare.wethers@utu.fi}

\author[0000-0002-7135-2842]{Clare F. Wethers}
\affiliation{Department of Space, Earth and Environment, Chalmers University of Technology, Onsala Space Observatory, 439 92 Onsala, Sweden}

\author{Nischal Acharya}
\affiliation{Donostia International Physics Centre (DIPC), Paseo Manuel de Lardizabal 4, E-20018 Donostia-San Sebastian, Spain}

\author{Roberto De Propris}
\affiliation{Finnish Centre for Astronomy with ESO (FINCA), Vesilinnantie 5, FI-20014 University of Turku, Finland}
\affiliation{Department of Physics and Astronomy, Vesilinnantie 5, FI-20014 University of Turku, Finland}

\author[0000-0003-0133-7644]{Jari Kotilainen}
\affiliation{Finnish Centre for Astronomy with ESO (FINCA), Vesilinnantie 5, FI-20014 University of Turku, Finland}
\affiliation{Department of Physics and Astronomy, Vesilinnantie 5, FI-20014 University of Turku, Finland}

\author{Ivan K. Baldry}
\affiliation{Astrophysics Research Institute, Liverpool John Moores University, 146 Brownlow Hill, Liverpool L3 5RF, UK}

\author[0000-0002-9796-1363]{Sarah Brough}
\affiliation{School of Physics, University of New South Wales, NSW 2052, Australia}

\author{Simon P. Driver}
\affiliation{ICRAR, The University of Western Australia, 35 Stirling Highway, Crawley WA 6009, Australia}
\affiliation{SUPA, School of Physics \& Astronomy, University of St Andrews, North Haugh, St Andrews, KY16 9SS, UK}

\author[0000-0002-6496-9414]{Alister W. Graham}
\affiliation{Center for Astrophysics and Supercomputing, Swinburne University of Technology, Hawthorn, VIC 3122, Australia}

\author[0000-0002-4884-6756]{Benne W. Holwerda}
\affiliation{Leiden Observatory, University of Leiden, Niels Bohrweg 2, 2333 CA Leiden, The Netherlands}

\author{Andrew M. Hopkins}
\affiliation{Australian Astronomical Optics, Macquarie University, 105 Delhi Rd, North Ryde, NSW 2113, Australia}

\author{Angel R. L{\'o}pez-S{\'a}nchez}
\affiliation{Australian Astronomical Optics, Macquarie University, 105 Delhi Rd, North Ryde, NSW 2113, Australia}
\affiliation{Department of Physics and Astronomy, Macquarie University, NSW 2109, Australia}
\affiliation{Macquarie University Research Centre for Astronomy, Astrophysics \& Astrophotonics, Sydney, NSW 2109, Australia}
\affiliation{ARC Centre of Excellence for All Sky Astrophysics in 3 Dimensions (ASTRO-3D), Australia}

\author{Jonathan Loveday}
\affiliation{Astronomy Centre, Department of Physics and Astronomy, University of Sussex, Falmer, Brighton BN1 9QH, UK}

\author{Steven Phillipps}
\affiliation{School of Physics, University of Bristol, Bristol BS8 1TL, UK}

\author{Kevin A. Pimbblet}
\affiliation{School of Physics and Monash Centre for Astrophysics, Monash University, Clayton, VIC 3800, Australia}
\affiliation{Department of Physics and Mathematics, University of Hull, Cottingham Road, Kingston-upon-Hull HU6 7RX, UK}

\author{Edward Taylor}
\affiliation{Center for Astrophysics and Supercomputing, Swinburne University of Technology, Hawthorn, VIC 3122, Australia}

\author{Lingyu Wang}
\affiliation{Institute for Computational Cosmology, Department of Physics, Durham University, Durham, DH1 3LE, UK}

\author{Angus H. Wright}
\affiliation{Astronomisches Institut, Ruhr-Universität Bochum, Universitätsstr. 150
44801 Bochum, Germany}



\begin{abstract}

Understanding the connection between nuclear activity and galaxy environment remains critical in constraining models of galaxy evolution. By exploiting extensive catalogued data from the Galaxy and Mass Assembly (GAMA) survey,  we identify a representative sample of 205 quasars at 0.1 $<$ z $<$ 0.35 and establish a comparison sample of galaxies, closely matched to the quasar sample in terms of both stellar mass and redshift. On scales $<$1 Mpc, the galaxy number counts and group membership of quasars appear entirely consistent with those of the matched galaxy sample. Despite this, we find that quasars are $\sim$1.5 times more likely to be classified as the group center, indicating a potential link between quasar activity and cold gas flows or galaxy interactions associated with rich group environments. On scales of $\sim$a few Mpc, the clustering strength of both samples are statistically consistent and beyond 10 Mpc we find no evidence that quasars trace large scale structures any more than the galaxy control sample. Both populations are found to prefer intermediate-density sheets and filaments to either very high- or very low- density environments. This weak dependence of quasar activity on galaxy environment supports a paradigm in which quasars represent a phase in the lifetime of all massive galaxies and in which secular processes and a group-centric location are the dominant trigger of quasars at low redshift. 

\end{abstract}



\section{Introduction} \label{sec:intro}

In the current paradigm of galaxy evolution, galaxies co-evolve alongside their central super-massive black hole. For decades, tight correlations have been observed between the black hole mass, M$_{\rm{BH}}$, and various properties of the parent galaxy bulge \citep{magorrian98,kormendy13,graham16}, which in turn have been shown to depend strongly on galaxy environment \citep{bahcall69}. In particular, early studies of the morphology-density relation \citep{oemler74,dressler80} found a convincing link between galaxy morphology and group- and cluster-scale environment, with star-forming disk-dominated galaxies typically residing in lower density environments than active ellipticals. Numerous studies have since supported this idea \citep[e.g.][]{lewis02,gomez03,balogh04,einasto05,gao05,gilmour07,porter08,skibba09,lietzen09,wang11}, finding actively star-forming galaxies to reside in under-dense environments. 

However, this relationship of galaxy environment with star-formation and morphology may not be universal. \cite{wijesinghe12}, for example, find no connection between environment and star formation among exclusively star-forming galaxies, seeing differences only in the environments of star-forming galaxies compared to passive galaxies. A similar dichotomy is observed in the slope of the M$_{\rm{BH}}$ - M$_{*,\rm{bulge}}$ relation, which appears much steeper for late type galaxies than for early type systems \citep[e.g.][]{davis18,davis19,sahu19}, leading to the idea that distinct blue and red sequences exist \citep{savorgnan16}. Furthermore, work by \cite{lietzen11} finds active and elliptical galaxies to appear more strongly influenced by environment than spiral galaxies. Understanding the link between galaxy properties and environment over a range of scales therefore remains an important test of galaxy evolutionary models.

On scales $\lesssim$ 1 Mpc, galaxy environment is sometimes used as an indirect tracer of galaxy interactions, with over-dense regions typically associated with a higher incidence of mergers. Such interactions may be responsible for triggering AGN activity \citep[e.g.][]{sanders88,barnes92,veilleux02,hopkins06}, funnelling gas into the central regions of the galaxy and fueling both star formation and accretion onto the black hole. Indeed, early studies \citep[e.g.][]{chu88,shanks88,disney95} found that luminous AGN, or \textit{quasars}, generally have very close companions and appear significantly more clustered than the general galaxy population out to $\sim$1 Mpc. Similarly, a more recent study by \cite{serber06} finds an overdensity in the environment of quasars compared to L* galaxies by up to a factor of three, with the strongest overdensities shown to exist around the most luminous quasars on scales $<$100 kpc. Several subsequent studies have also supported these findings, detecting an enhancement in the merger fraction of luminous (L$_{\rm{bol}}$ $>$10$^{45}$ ergs$^{-1}$), high redshift quasars \citep[e.g.][]{kocevski11,treister12} and leading to the idea that galaxy interactions may be required to trigger these systems.

At lower redshifts and among lower luminosity quasar populations, the connection between nuclear activity and galaxy interactions is less clear, with a number of studies finding quasar environments to be consistent with those of the general galaxy population. A study by \cite{karhunen14} for example, finds no difference in the number density within a projected 1 Mpc radius of $z<$0.5 quasars compared to inactive galaxies at the same redshift, matched in luminosity. Likewise, \cite{coldwell06} find the local environments of $z<$0.2 SDSS quasars to be similar to typical galaxies. This seemingly weak dependence of quasar activity on local environment contradicts the high redshift paradigm in which major mergers are required to trigger nuclear activity. The similar local environments of quasars and typical galaxies at low redshifts may alternatively support the triggering of quasars via secular processes, such as stochastic gas accretion, minor mergers and bar instabilities. While these triggering mechanisms are typically associated with low-luminosity quasars, a handful of studies have suggested that such secular processes my be sufficient in triggering, and subsequently fuelling, even the most luminous quasars at low redshift \citep[e.g.][]{cisternas10, villforth14}.

On larger scales of $\sim$a few Mpc, early studies found quasars to preferentially reside in environmental overdensities comparable to galaxy groups or poor clusters \citep[e.g.][]{stockton78,yee84}, but more recent work by \citep{zhang13} suggests that clustering strength strongly evolves with both M$_{\rm{BH}}$ and redshift out to $z$=2. Similarly, while a study by \cite{sochting02} found $z<$0.4 quasars to trace the large-scale ($>$10 Mpc projected distance) structures populated by galaxy clusters, several newer studies find no such correlation. Both \cite{miller03} and \cite{villforth12} for example, demonstrate nuclear activity to be independent of the projected $>$10 Mpc environment, concluding that quasars show no preference towards either very high- or low-density environments over large scales. 


With the advent of large field surveys such as the Sloan Digital Sky Survey \citep[SDSS: ][]{blanton17} and the 2 degree Field Galaxy Redshift Survey \citep[2dFGRS: ][]{colless01}, it has become possible to study the environments of ever-larger statistical quasar samples at low redshift. Indeed, several studies have taken advantage of this \citep[e.g.][]{croom04,serber06,zhang13}, yet selecting a robust galaxy control sample with which to compare the results of such studies remains challenging. Over the last decade, the Galaxy and Mass Assembly Survey \citep[GAMA: ][]{liske15,baldry18} has opened the door to not only studying large quasar samples but also to selecting large galaxy comparison samples, based on a range of properties. For the first time, GAMA has provided information on the group, cluster and large scale environments of $\sim$300,000 galaxies at low redshift. The extensive coverage of GAMA means the properties of quasars and inactive galaxies can be directly compared, as their derived properties will be subject to the same set of biases and limitations. Throughout this work, we exploit the large survey area and high completeness of GAMA to investigate the the environments of z$<$0.3 quasars out to a projected distance of $>$10 Mpc and compare them to the underlying galaxy population, matched in both redshift and stellar mass. In this way, we seek to test the idea that quasars are commonplace as a phase in the lifetime of galaxies and comment on the likely triggering mechanisms for quasar activity at low redshift based on their environmental properties.

This paper represents the first in a series of papers exploring the properties of quasars in GAMA and is structured as follows. Section~\ref{sec:data} details the quasar sample considered in this work, along with the matched galaxy comparison sample. In Section~\ref{sec:results}, we explore the environments of quasars in GAMA over a range of scales from $\sim$100 kpc out to $>$10 Mpc. Our key results are summarised in Section~\ref{sec:conclusions}. Throughout this paper, we assume a flat $\Lambda$CDM cosmology with $H_{0}$ = 70 km s$^{-1}$ Mpc$^{-1}$, $\Omega_{M}$ = 0.3 and $\Omega_{\Lambda}$ = 0.7. Unless otherwise specified, all quoted magnitudes are given in the AB system.

\section{Data} \label{sec:data}

Throughout this paper, we make use of proprietary data from the latest internal data release of GAMA. GAMA is a wide-field spectroscopic survey, observing some $\sim$300k galaxies using the 2dF multi-fiber instrument \citep{lewis02} in combination with the AAOmega spectrograph \citep{saunders04,smith04,sharp06} on the Anglo-Australian Telescope (AAT). The 2dF instrument, which is installed at the prime focus of the AAT, positions $>$400 optical fibers to provide a 2 degree (diameter) field of view with a fibre-positioning accuracy of 0.3 arcseconds. The full GAMA survey, carried out between 2008 February and 2014 September, covers $\sim$286 deg$^2$ of the southern sky over five fields, each covering $\sim$60 deg$^2$. In this work, we consider only the three equatorial survey regions (G09, G12 and G15), over which the survey is most complete ($>$98 per cent to m$_r$=19.8) and for which the most extensive ancillary data is available. Table~\ref{tab:gama_fields} presents the sky coverage and depth for each of the equatorial fields in GAMA. The photometric input catalogue for these regions is fundamentally based on the Sloan Digital Sky Survey \citep[SDSS: ][]{york00} and is described in detail in \cite{baldry10}. Details of the redshift measurements and spectroscopic pipeline used in the survey are available in \cite{hopkins13} and \cite{liske15}.

\begin{table}
    \centering
    \begin{tabular}{c|c|c|c}
    \hline
    Region    &  RA (deg) &  Dec (deg) & Depth ($r_{\rm{AB}}$) \\
    \hline
    G09 & 129.0 - 141.0 & -2.0 - +3.0 & $<$19.8 \\
    G12 & 174.0 - 186.0 & -3.0 - +2.0 & $<$19.8 \\
    G15 & 211.5 - 223.5 & -2.0 - +3.0 & $<$19.8 \\
    \hline
    \end{tabular}
    \caption{Sky coverage of the three equatorial GAMA survey regions considered throughout this paper (G09, G25 and G15), along with the $r$-band survey depth of each field.}
    \label{tab:gama_fields}
\end{table}

\subsection{Quasars in GAMA} \label{sec:gama_qsos}

Quasar targets are initially selected from the fourth version of the Large Quasar Astrometric Catalogue (LQAC-4), which identifies a near-complete sample of $>$400,000 Type-I quasars \citep{gattano18}, spectroscopically confirmed as such from their broad optical line emission. LQAC-4 is the most homogeneous and complete quasar catalogue to date, cross-matching 12 independent quasar surveys, alongside the Veron-Cetty \& Veron quasar catalogue, to provide $ubvgrizJK$-band photometry and radio fluxes at 1.4GHz, 2.3GHz, 5.0GHz, 8.4GHz and 24GHz, along with spectroscopic redshifts. Initially, we isolate all quasars in LQAC-4 overlapping the three equatorial regions of the GAMA survey and apply a redshift cut of 0.1$<z<$0.35, corresponding to the range in redshift over which GAMA is most complete. Targets are further required to have an $r$-band magnitude, m$_{r}$, brighter than the GAMA survey depth. In order to minimise potential selection biases and ensure our sample is representative of the low-redshift quasar population, we do not impose any additional selection criteria, recovering an initial sample of 230 quasars.

\begin{figure}
\includegraphics[trim= 25 5 30 15 ,clip,width=.5\textwidth]{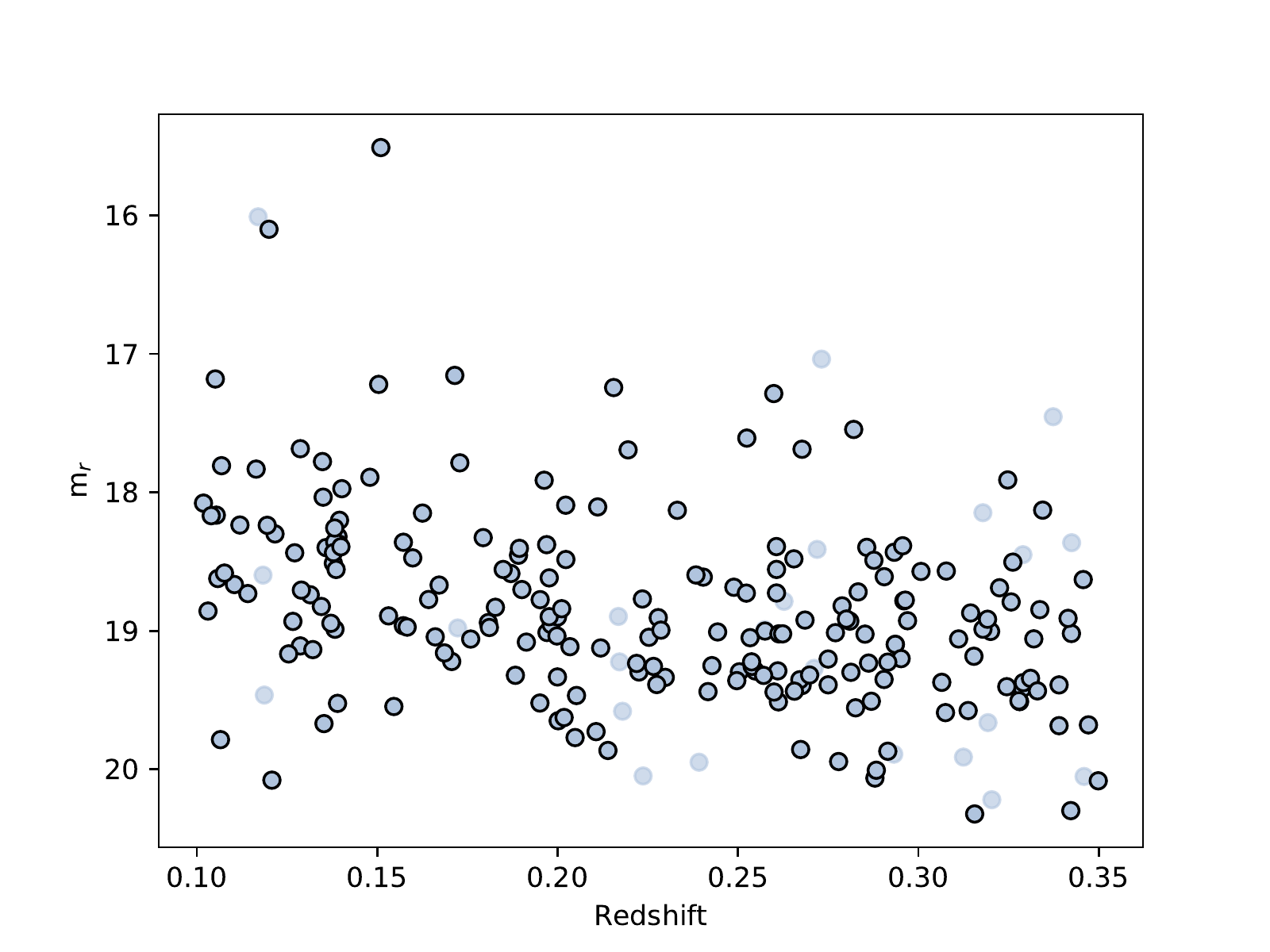}
\caption{$r$-band apparent magnitude (AB) vs. redshift for all MAGPHYS quasar targets in GAMA (\textit{black circles}) compared to the full sample of quasars in LQAC-4 over the same redshift range brighter than the GAMA magnitude limit (\textit{blue dots}).}
\label{fig:sample_bias}
\end{figure}

Although GAMA is not specifically targeted to find quasars and is biased against bright point sources that fail the star/ galaxy separation criteria \citep{baldry10}, we highlight that quasars at 0.1$<z<$0.35 typically appear extended, as light from the host galaxy can be spatially resolved. Indeed, positional cross-matching of the 230 LQAC-4 quasars with GAMA ($<$5 arcsec) returns 205 quasars, meaning just 25 of the 230 quasars in LQAC-4 ($\sim$10 per cent) are missed by GAMA, potentially due to this point source exclusion. To ensure this does not bias our sample, we plot the 230 LQAC-4 quasars as a function of both redshift and m$_{r}$, highlighting those with counterparts in GAMA (Fig.~\ref{fig:sample_bias}). The resulting GAMA quasar sample covers the full range of redshifts and m$_{r}$ of the quasars in LQAC-4, demonstrating that the point source exclusion of GAMA does not bias the quasar sample. Rather, the subset of 205 quasars in GAMA, which form the basis of this work, are representative of the general quasar population at 0.1$<z<$0.35.

\begin{figure}
\includegraphics[trim= 0 5 30 15 ,clip,width=.5\textwidth]{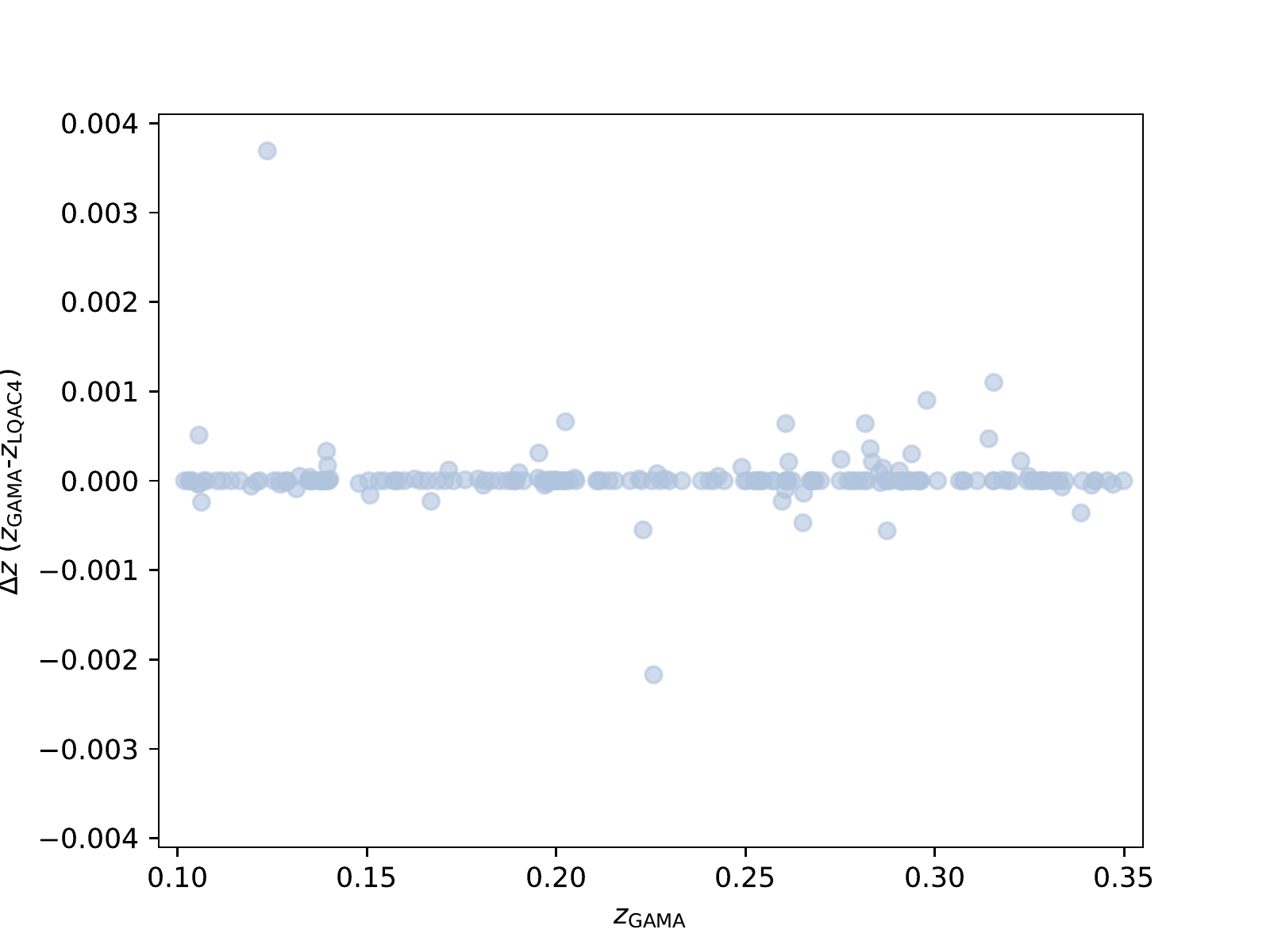}
\caption{$\Delta z$($z_{\rm{GAMA}}$ - $z_{\rm{LQAC4}}$) as a function of $z_{\rm{GAMA}}$ for the 205 quasars in our sample. $\Delta z<$ 0.004 in all cases.}
\label{fig:z_bias}
\end{figure}

In addition to the 205 quasars identified in this manner, two further quasars are identified in GAMA, which have LQAC-4 redshifts lying outside the redshift range of our sample and differ from those in GAMA by $\Delta z>$ 0.80. These two targets are therefore not included in our sample. To ensure this is not an issue for the remainder of the quasar sample, we compare the redshift estimates derived from both surveys (LQAC-4 and GAMA) across the full quasar sample (Fig.~\ref{fig:z_bias}), finding a near-perfect match ($\Delta z<$0.004) between the two sets of derived redshifts. While we refer to the GAMA redshifts throughout this work, we highlight that instead choosing to use the LQAC-4 redshifts would make no difference to the results of the study.

\begin{figure}
\includegraphics[trim= 0 0 10 0 ,clip,width=.5\textwidth]{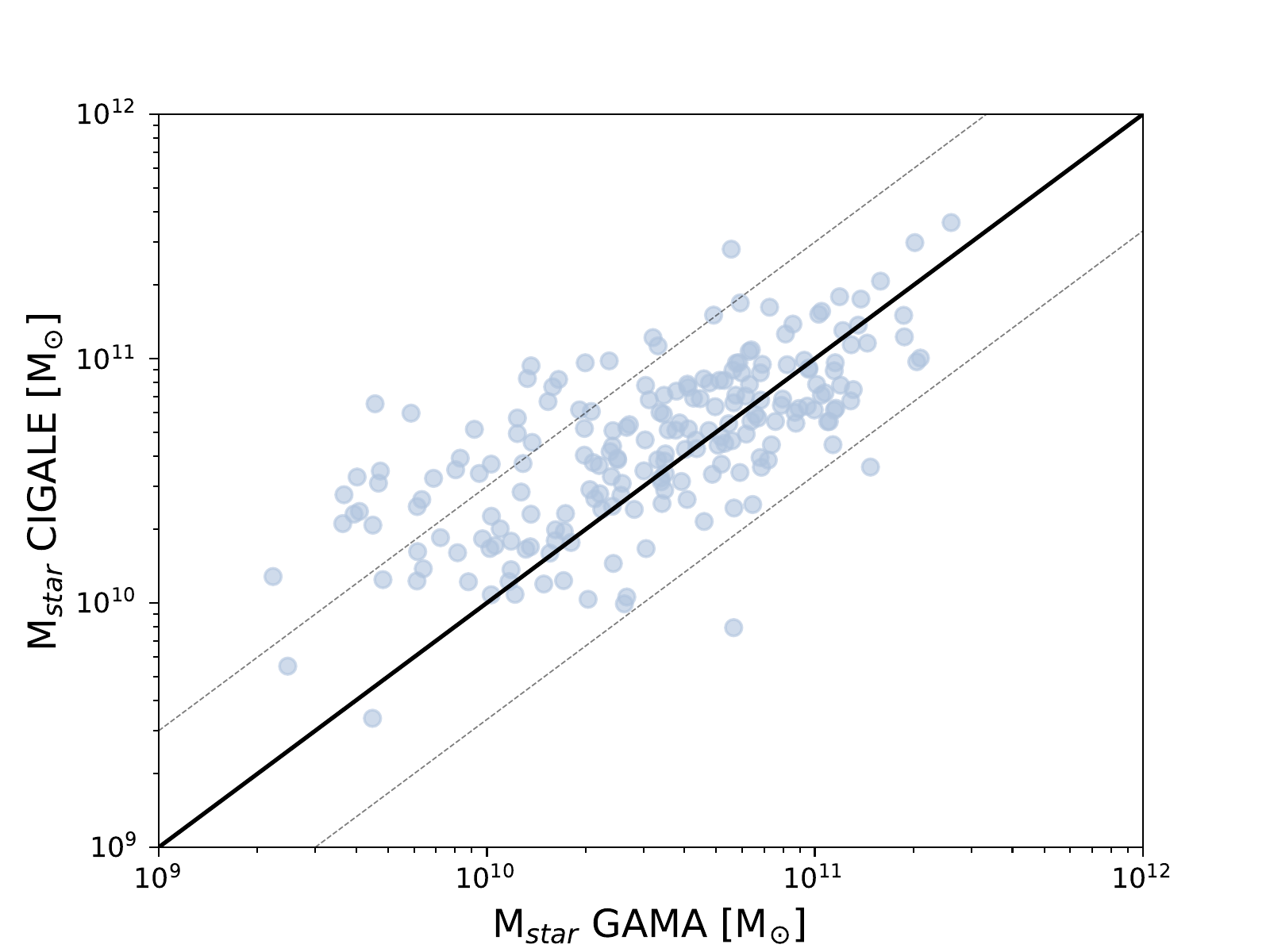}
\caption{Comparison of the stellar mass derived for our quasar sample in GAMA and from the \textsc{cigale} SED fitting. Dotted lines denote regions of $\pm$0.2 dex, within which 82.93 per cent of the sample lie.}
\label{fig:mass_comp}
\end{figure}

Each of the 205 quasars in our sample have optical spectra from either the sixteenth data release (DR16) of SDSS or newer observations from the AAT, which confirm them to be Type-I quasars at redshifts 0.1$<z<$0.35. In addition, all targets have GAMA-derived stellar mass estimates \citep[\textsc{StellarMassesv20}: ][]{taylor11}, based on the fitting of spectral energy distributions (SEDs) to the 21-band panchromatic photometry (\textsc{LambdaPhotometryv03}: \citeauthor{wright17} \citeyear{wright17}; \textsc{PanchromaticPhotomv01}: \citeauthor{driver16} \citeyear{driver16}) using the MAGPHYS \citep{da11} code \citep[\textsc{MagPhysv06}:][]{driver18}. However, due to the exclusion of bright point sources in GAMA, this SED fitting routine does not consider any contribution from the quasar and thus may overestimate the stellar mass of our quasar sample. To test the extent of this potential bias, we independently fit the same GAMA photometry using version 20 of the Code Investigating GALaxy Emission (CIGALE: \citealt{noll09,burgarella15,boquien19}). Unlike the fitting used by GAMA, our SED combines a quasar template with the stellar population model \citep{bruzual03}, accounting for both nebular \citep{inoue11} and dust emission \citep{draine14} attenuated by a power-law suitable for local star-forming galaxies. In general, the quasar contribution to the overall flux is found to be relatively small, accounting for $\geq$50 per cent of the emission in $<$30 per cent of our sample, although we acknowledge that accurately measuring the extent of quasar contamination remains a complex issue well beyond the scope of this work. Nevertheless, due to this relatively small level of quasar contamination in the sample we derive stellar masses consistent to within 0.2 dex of those derived in GAMA for 82.93 per cent of our sample (Fig.~\ref{fig:mass_comp}). Although we opt to use the CIGALE stellar mass estimates derived for the quasar hosts to select our matched galaxy sample (section~\ref{sec:matched_gals}), we therefore highlight that using the GAMA stellar masses would produce a similar mass distribution from which to select the comparison sample.

\subsection{Matched Galaxy Sample} \label{sec:matched_gals}

From the $\sim$300k galaxies observed by GAMA, $\sim$120k are included in \textsc{MagPhysv06} \citep{driver16}, which provides information on the stellar populations and ISM of GAMA galaxies over the three equatorial survey regions. From these $\sim$120k galaxies we remove all quasar hosts to create the pool from which to sample the matched comparison galaxies. We note that the the stellar mass estimates in GAMA are sufficient here, as there is no quasar component to be accounted for. We therefore do not derive independent mass estimates for this population. Instead, we select galaxy comparison samples based on their GAMA-derived stellar masses, matched to the independent (CIGALE) masses of the quasar sample (see section~\ref{sec:gama_qsos}), which account for the additional quasar component. 

\begin{figure*}
	\centering
	\subfigure[]{\includegraphics[trim=0 0 0 0,clip,width=0.5\textwidth]{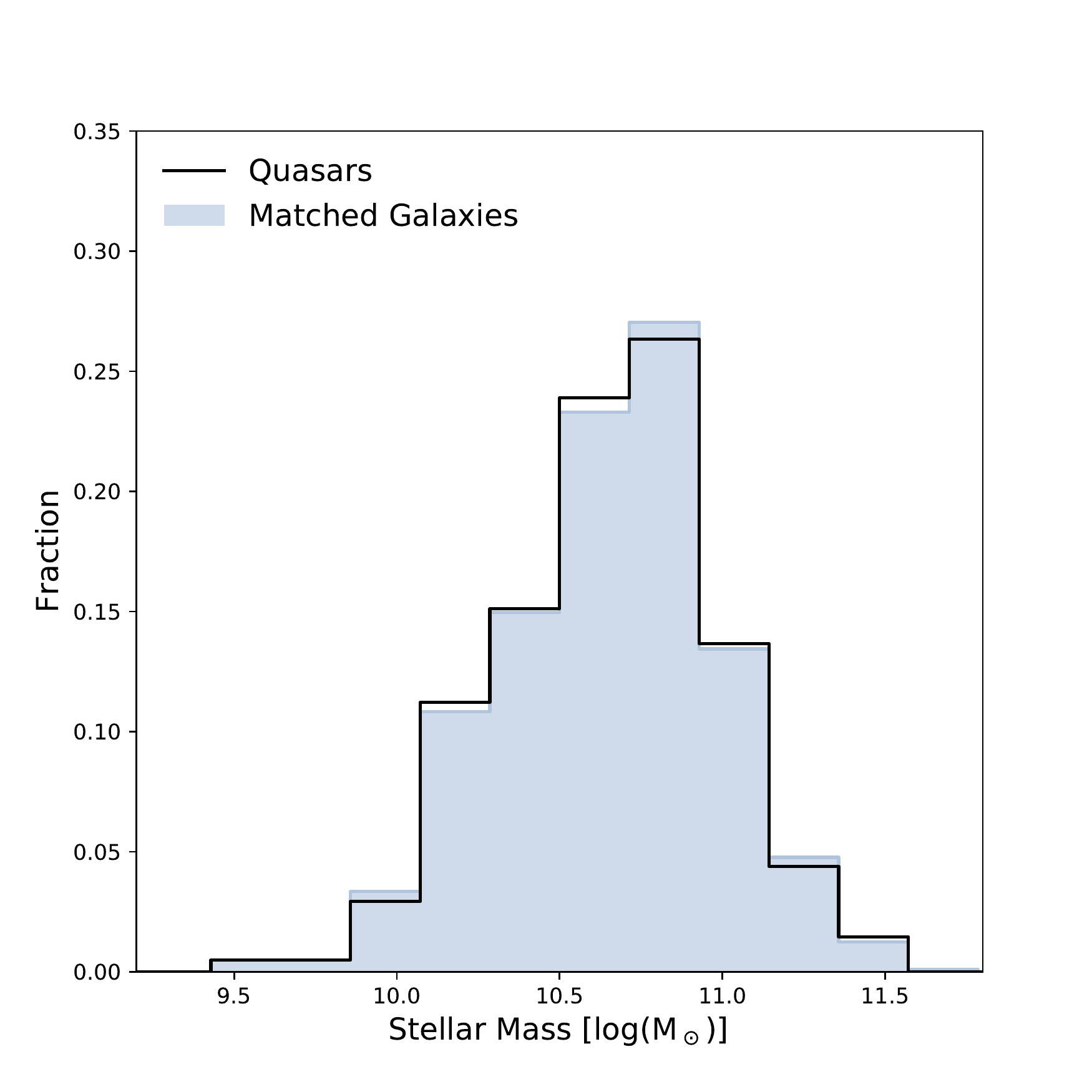}}\hfill
	\subfigure[]{\includegraphics[trim=0 0 0 0,clip,width=0.5\textwidth]{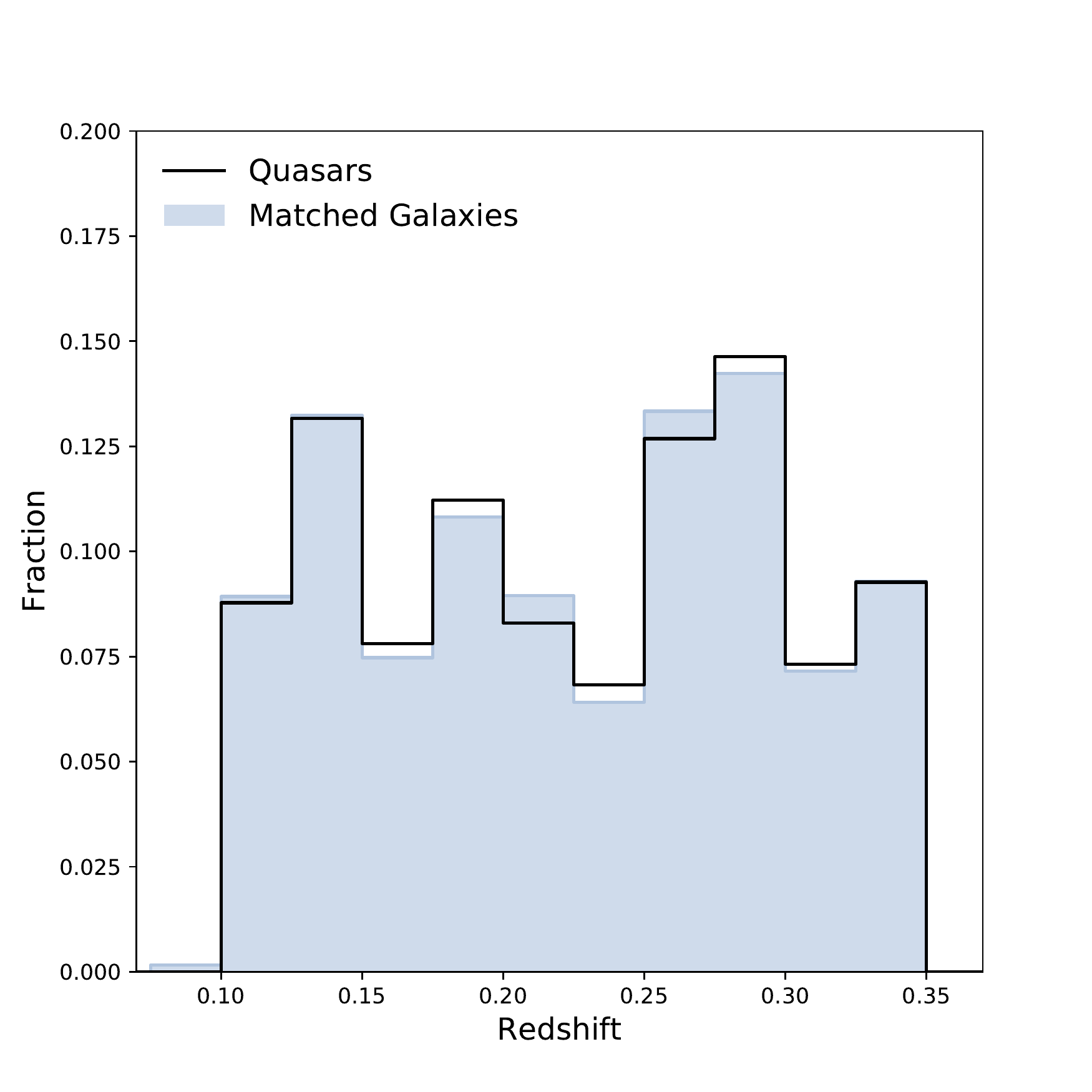}}\hfill	
\caption{Distribution of stellar masses (a) and redshifts (b) for quasars in GAMA (\textit{black}) and the matched galaxy sample (\textit{blue}).}
\label{fig:sampleAll}
\end{figure*}

In order to select galaxies closely matched in redshift, $z$, and stellar mass, M$_{*}$, to the quasar hosts in GAMA, we then define a quantity, $\Delta$C;

\begin{equation}
    \Delta C = \left(\frac{ z-z_{\rm{QSO}}}{0.01}\right)^2 + \left(\frac{ M_*-M_{*,\rm{QSO}}}{0.1}\right)^2,
\end{equation}

accounting for a tolerance of 0.01 and 0.1 dex in $z$ and M$_{*}$ respectively. $\Delta$C is calculated for every GAMA galaxy in reference to each quasar in turn. For each quasar, we select the 100 galaxies for which the lowest values of $\Delta$C are derived. To create a single realisation of the matched galaxy sample, one galaxy is selected at random from each set and removed from the selection pool. This results in a single comparison sample of 205 galaxies. The process is then repeated to obtain 100 realisations of this galaxy sample, each closely matched to the GAMA quasars in both $z$ and M$_{*}$. The distribution of M$_{*}$ and $z$ across the quasar sample and the final pool of 205$\times$100 galaxies in GAMA are shown in Fig.~\ref{fig:sampleAll}. Obtaining closely matched comparison samples in this way is vital in order to eliminate any potential environmental biases arising from differences in $z$ or M$_{*}$ and allows direct comparisons to be made between the properties of the two populations.

\section{Results and Discussion} \label{sec:results}

\subsection{Local Environment} \label{sec:local_env}

On scales $\lesssim$1 Mpc, galaxy environment is often used as an indirect tracer of interactions, with overdense local environments typically indicating a higher frequency of galaxy interactions. Characterizing quasar environments on these scales is therefore critical in understanding the role that galaxy interactions play in triggering quasar activity. In particular, the environments of quasars on scales of $\sim$100 kpc have been shown to exhibit the strongest overdensities, with \cite{serber06} finding quasars to reside in environments a factor of 1.4 times more dense than L$^{*}$ galaxies on these scales. This overdensity is postulated to decrease monotonically with the scale of the environment, decreasing to unity $\gtrsim$1 Mpc. Owing to the extensive coverage of the GAMA survey, we are now able to directly compare the environments of quasars on scales of $\sim$100 kpc with a comprehensive sample of galaxies, closely matched in both stellar mass and redshift.

To explore the local environments of quasars in GAMA, the sky positions of our quasar targets are cross-matched with version 27 of the spectroscopic catalogue (\textsc{SpecObjv27}) from the internal data release of GAMA.  This catalogue contains all spectroscopic GAMA sources over the three equatorial survey regions reaching a $\sim$98 per cent completeness down to m$_{r}$=19.8. All sources lying within a projected separation of 100 kpc and a distance of $\Delta V<$1000 kms$^{-1}$ in velocity space of each quasar target are counted, excluding the target itself. Here, the projected distance of 100 kpc has been chosen to match the area used in the work of \cite{serber06}, who find this separation to host the strongest environmental overdensities around quasars. Likewise, $\Delta V<$1000 kms$^{-1}$ has been selected in accordance with several other works \citep[e.g.][]{muldrew12,shattow13,moon19}. Based on the number counts over these parameters, we recover an average neighbor count for the GAMA quasars of $\bar{n}_{QSO}$ = 0.22 $\pm$ 0.03, with an uncertainty, S$_{n}$, given by  

\begin{equation}
    S_{n} = \sqrt{\frac{\bar{n}}{N}},
\end{equation}

where $\bar{n}$ is the average neighbor count and $N$ denotes the sample size (i.e. $N$=205 for the GAMA quasar sample). 

\begin{figure}
\includegraphics[trim= 15 5 30 15 ,clip,width=.5\textwidth]{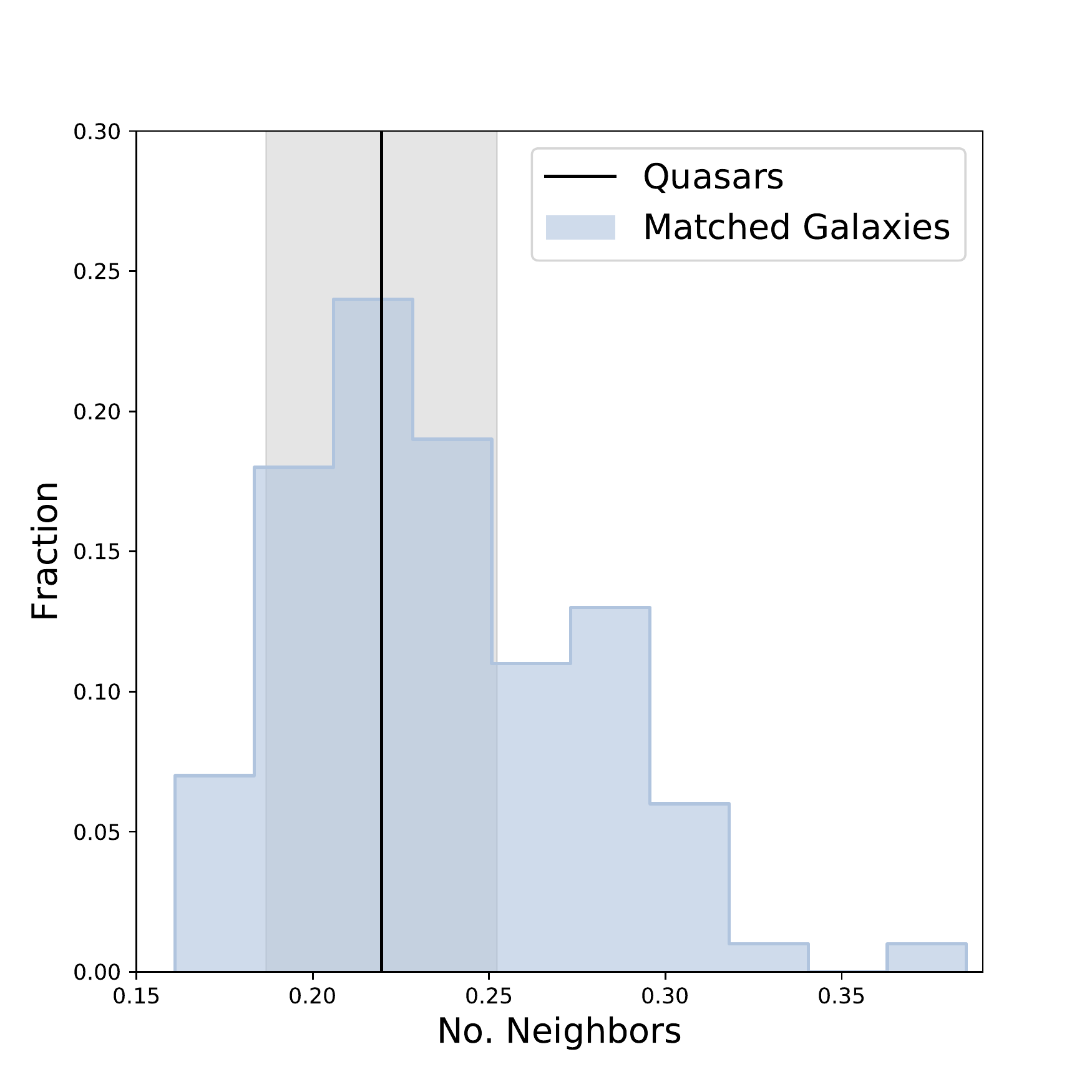}
\caption{Average number of neighbors within 100 kpc and $\Delta$V$<$1000 kms$^{-1}$ of the GAMA quasars (\textit{black line}) compared to the distribution of average neighbor counts for the 100 realisations of the matched galaxy sample (\textit{blue histogram}).}
\label{fig:neighbors}
\end{figure}

Similarly, we count the number of sources within 100 kpc and $\Delta V<$1000 kms$^{-1}$ of our matched galaxy sample. In this case, the average number of neighboring galaxies is calculated for each of the 100 realisations of the matched galaxy sample to create a \textit{distribution} of average neighbor counts, which can be directly compared to that of the quasar sample. Based on this distribution, we derive $\bar{n}_{GAL}$ = 0.24 $\pm$ 0.04, where the quoted uncertainty denotes the standard deviation (1$\sigma$) of the distribution (Fig.~\ref{fig:neighbors}). Based on Fig.~\ref{fig:neighbors}, we conclude the local environments of quasars to be similar to those of the matched galaxy sample. \textit{The average neighbor counts of each population are entirely consistent to within the quoted uncertainties, indicating no difference in the $<$100 kpc environment of quasars and galaxies matched in stellar mass at low redshift (0.1$< z<$0.35).} 

\begin{figure*}
\includegraphics[trim= 0 0 0 0 ,clip,width=\textwidth]{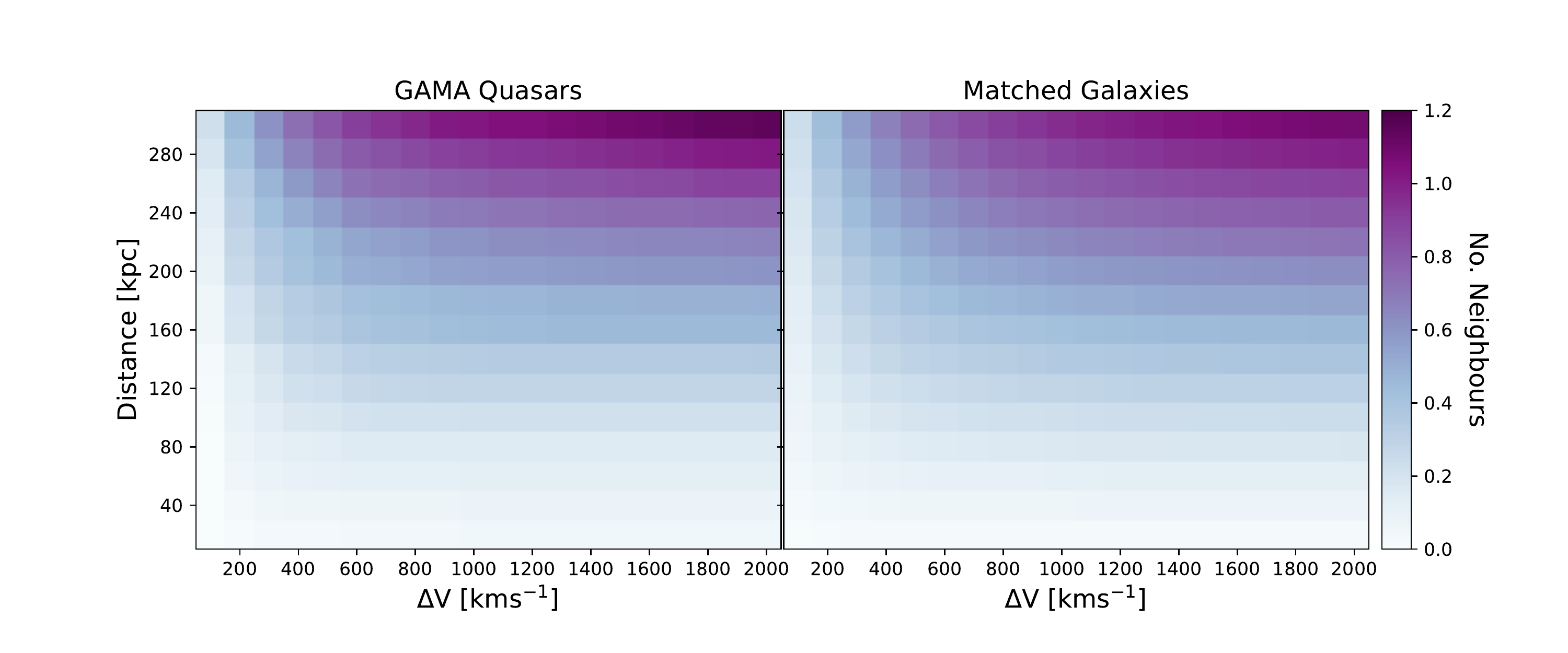}
\caption{Heat maps showing the median average number of neighbours within radii ranging 20 to 300 kpc and $\Delta$V ranging 100 to 2000 kms$^{-1}$ of the GAMA quasars (\textit{left}) and the matched galaxy sample (\textit{right}).}
\label{fig:heatmap}
\end{figure*}

\begin{figure}
\includegraphics[trim= 90 0 20 10 ,clip,width=.5\textwidth]{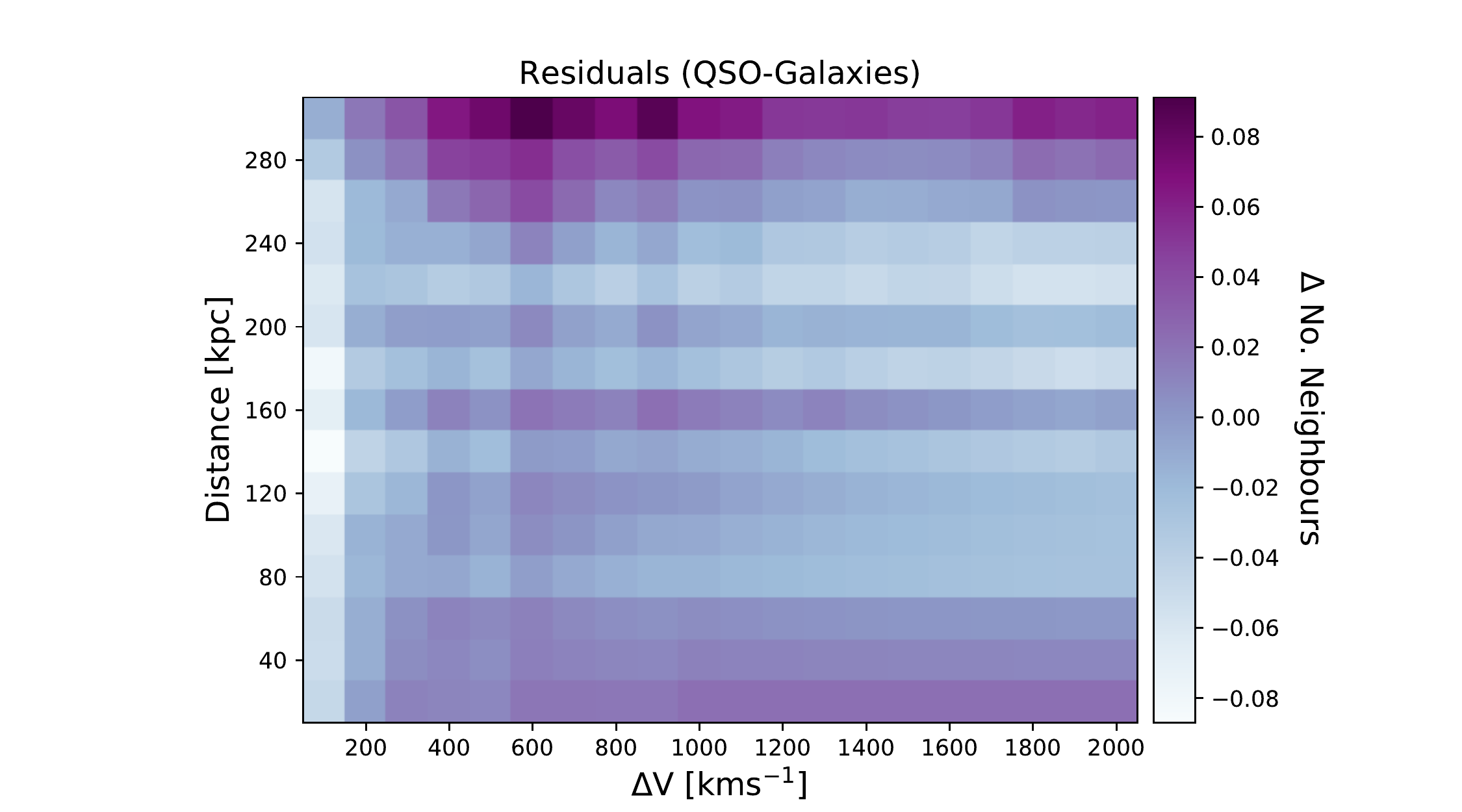}
\caption{Quasar - galaxy residuals for the heat maps in Fig.~\ref{fig:heatmap}.}
\label{fig:residuals}
\end{figure}

To test the extent of the apparent similarity in the local environments of quasars and the underlying galaxy population, we extend the above study to cover a range of physical scales and $\Delta$V. To this end, we obtain number counts for both our quasar and matched galaxy samples within radii and $\Delta$V ranging 20 to 300 kpc and 100 to 2000 kms$^{-1}$ respectively. The median number of sources within each radii and $\Delta$V are then calculated. The resulting average number counts for each population (quasars and matched galaxies), along with the quasar - matched galaxy residuals, are given in Figs.~\ref{fig:heatmap} and ~\ref{fig:residuals} respectively. Although both the quasar and galaxy maps appear almost identical (Fig.~\ref{fig:heatmap}), the residuals (Fig.~\ref{fig:residuals}) indicate a slight enhancement in the neighbour counts around quasars both at the closest ($\lesssim$60 kpc) and furthest ($\gtrsim$260 kpc) separations. We therefore suggest that any environmental overdensities around quasars likely occur on these scales, with little difference in the environments of the two populations over scales $\sim$70-250 kpc. We note however, that the difference in the average number of neighbours for each population remains small ($<$0.1), even in the most extreme cases. There is therefore little indication that the local environments of quasars are different to those of the underlying galaxy population at z$<$0.35.

Our results are in direct agreement with \cite{karhunen14}, who also find no statistical differences between the local environments of quasars and inactive galaxies at $z<$0.5, based on the projected number counts of $\sim$300 quasars and inactive galaxies matched in luminosity and redshift. This similarity in the local environments of the quasar and matched galaxy populations, both in \cite{karhunen14} and in our study, indicates a weak dependence of nuclear activity on local environment, which may in turn favour the secular triggering of quasars at low redshift. However, due to the 2 arcsec convolution limit of SDSS, corresponding to a physical scale of $\sim$10 kpc at the upper redshift limit of our sample, we note that we cannot rule out the possibility of close merger pairs within $\lesssim$10 kpc. While we do not find any evidence to suggest quasars exist in overdense environments on scales $<$100 kpc, we therefore cannot rule out triggering of quasars via close pair mergers.

\subsection{Group Environments} \label{sec:group_env}

On larger sub-Mpc (group) scales, quasars have been associated with enhancements in the spatial distribution of galaxies \citep{bahcall69}, typically residing in small-to-moderate group environments in both the local universe \citep[e.g.][]{bahcall91,fisher96,mclure01,karhunen14} and at higher redshifts \citep{hennawi06,stott20}. However, while some high redshift quasars appear to reside in dense environments, several others do not (see, e.g., \citealt{habouzit2019}, their figure 1). Given the association of rich group-scale environments with galaxy interactions and, by extension, the onset of nuclear activity following major mergers, understanding the group-scale environments of quasars remains an important test of their triggering and fuelling mechanisms.

\begin{figure}
\includegraphics[trim= 25 5 30 15 ,clip,width=.5\textwidth]{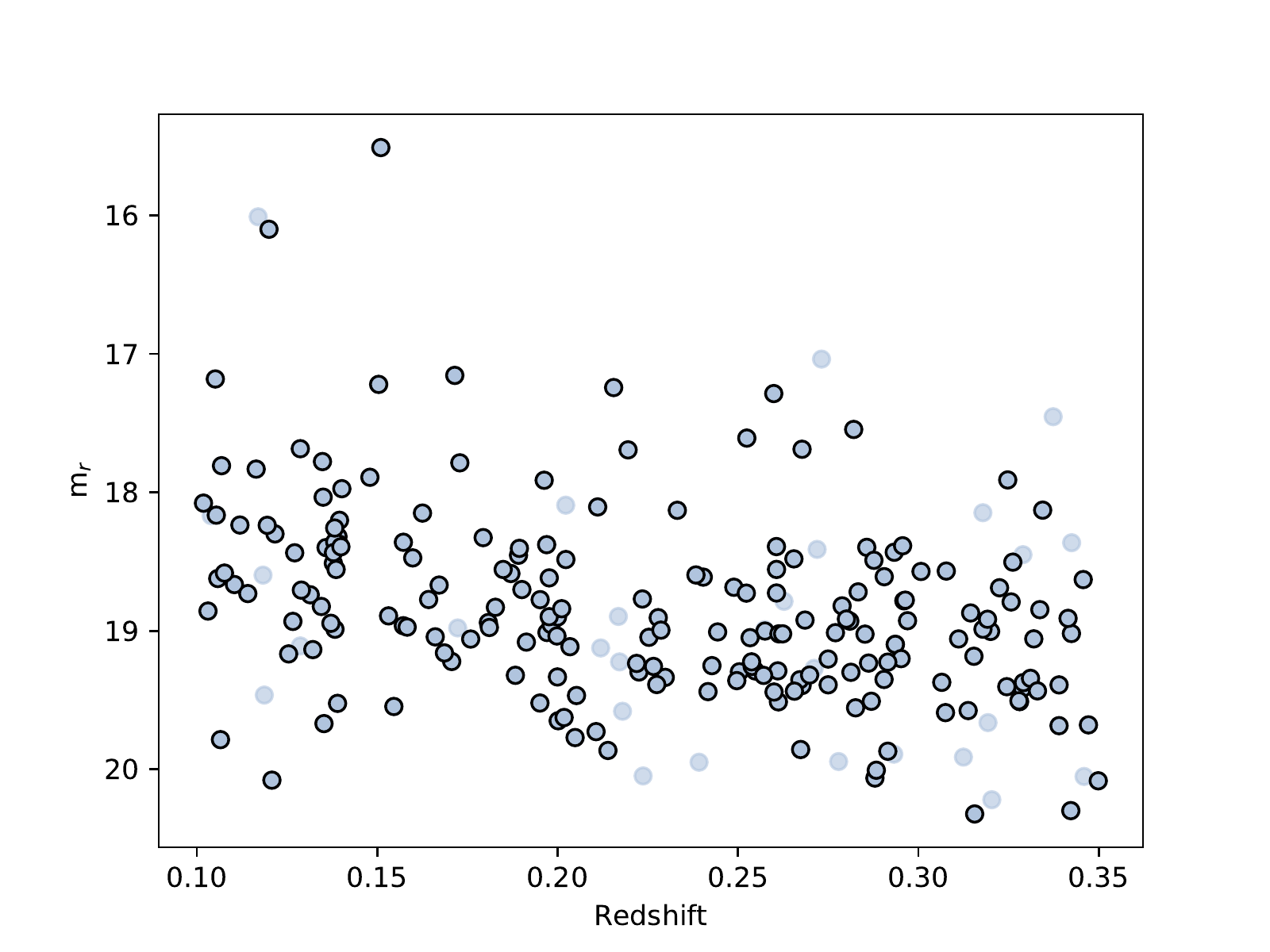}
\caption{$r$-band apparent magnitude (AB) vs. redshift for all quasar targets with group information in GAMA (\textit{black circles}) compared to the full sample of quasars in LQAC-4 over the same redshift range brighter than the GAMA magnitude limit (\textit{blue dots}).}
\label{fig:sample_bias_group}
\end{figure}

Of the 205 quasars in the GAMA sample, 200 are included in the \textsc{GroupFindingv10} catalog, which details their group properties, including whether or not they exist in a group and their position within that group with regards to the central galaxy \citep{robotham11}. The distribution of the subset of 200 quasars in terms of redshift and $r$-band magnitude is given in Fig.~\ref{fig:sample_bias_group}. We note that the 200 targets continue to provide a representative sample of the parent population of quasars taken from LQAC-4, covering the full range of $z$ and M$_{r}$ without any obvious bias.

\begin{figure*}
	\centering
	\subfigure[]{\includegraphics[trim=0 0 0 0,clip,width=0.5\textwidth]{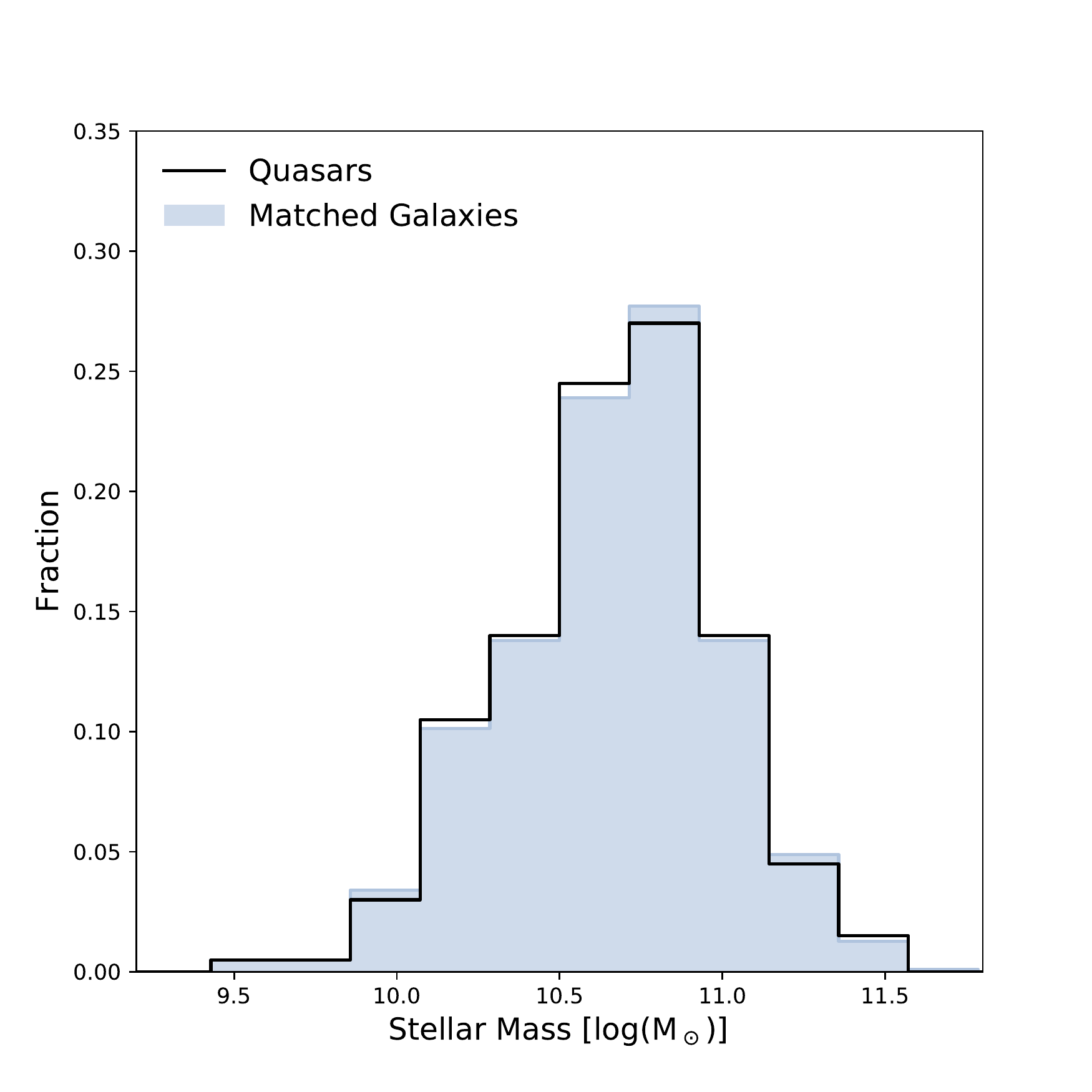}}\hfill
	\subfigure[]{\includegraphics[trim=0 0 0 0,clip,width=0.5\textwidth]{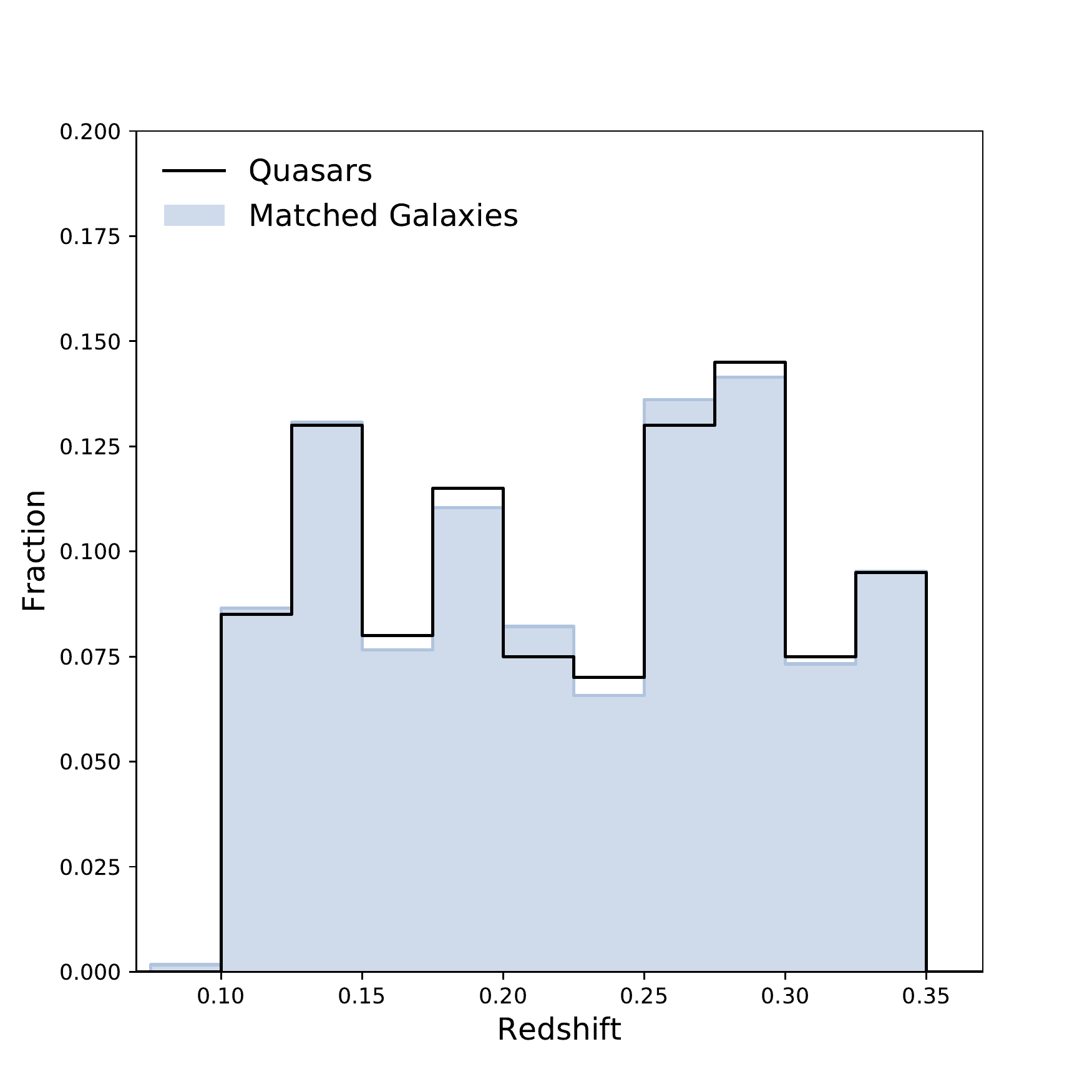}}\hfill	
\caption{Distribution of stellar masses (a) and redshifts (b) for quasars in GAMA with group information (\textit{black}) and the matched galaxy sample for which this information is also available (\textit{blue}).}
\label{fig:sample_group}
\end{figure*}

Although the reduction in sample size from our initial GAMA quasar sample is small (excluding just five targets), we nevertheless account for this size reduction in our matched galaxy sample. As such, we re-sample the 100 realisations of our galaxy sample, each consisting of 200 galaxies matched in stellar mass and redshift to our quasar sample. The redshift and stellar mass distributions of the resulting 100 $\times$ 200 galaxies compared to that of our quasar sample are shown in Fig.~\ref{fig:sample_group}. 

\subsubsection{Galaxies in Groups} \label{sec:group_gals}

\begin{figure}
\includegraphics[trim= 15 5 30 15 ,clip,width=.5\textwidth]{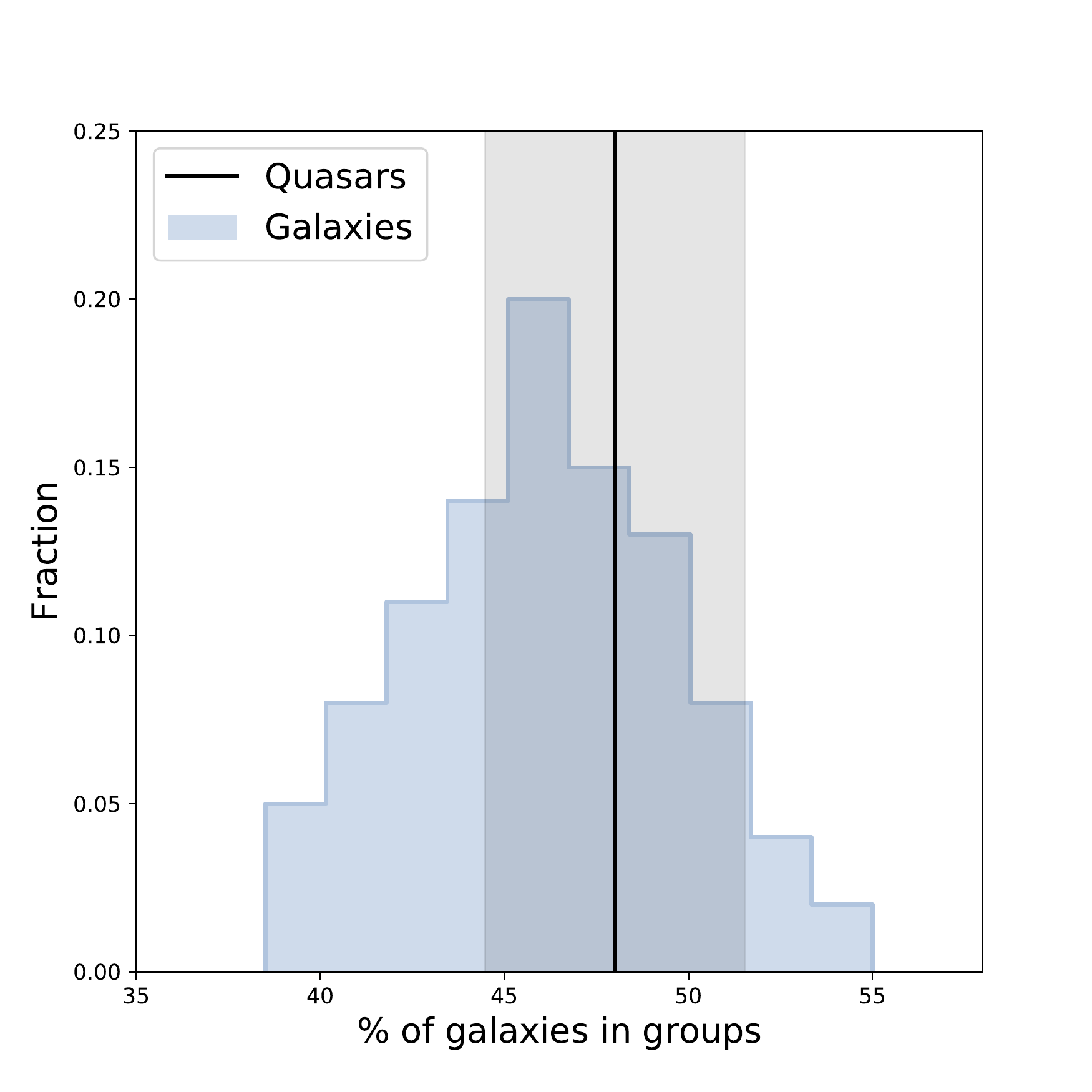}
\caption{Fraction of quasars in groups (\textit{black line}) compared to the distribution of group fractions across the 100 realizations of our matched galaxy sample (\textit{blue histogram}). Shaded regions denote the uncertainty on the quasar group fraction, derived in Eqn.~\ref{eqn:error}.}
\label{fig:group}
\end{figure}

The group properties of galaxies in GAMA are classified using a friends-of-friends (FoF) algorithm, which determines whether galaxies are associated with one another based on both their projected and co-moving separations \citep{robotham11}. The GAMA FoF grouping algorithm has been extensively tested on a set of mock GAMA light cones, derived from semi-analytic $\Lambda$CDM N-body simulations. The algorithm has been demonstrated to recover galaxy group properties with an accuracy of $>$80 per 
cent and is shown to be robust to the effects of outliers. Full details of the FoF algorithm used here can be found in \cite{robotham11}. This FoF algorithm classifies 96 of the 200 quasars as group galaxies, corresponding to 48.00 $\pm$ 3.53 per cent of the sample. The quoted uncertainty is estimated as the standard error, $S$, such that;

 \begin{equation}
    S = \sqrt{\frac{p (1-p)}{N}},
    \label{eqn:error}
\end{equation}

where $N$ denotes the number of quasars in the sample and $p$ is the fraction of quasars with a given property, in this case the fraction of quasars existing in a group. Similarly, we calculate the number of group galaxies in each of the 100 realisations of our matched galaxy sample, finding an average group fraction of 46.07 $\pm$ 3.57 per cent, where the uncertainty here denotes the standard deviation of the resulting distribution (Fig.~\ref{fig:group}). 
 
To test whether the difference between the group fraction of the two samples is significant, we perform a statistical $p$-value test to determine how likely it is that the two populations are drawn from the same underlying distribution, while accounting for the different sample sizes. To this end, we calculate the Z-statistic for each sample, i.e.

\begin{equation}
    Z = \frac { \overline{p}_{1} - \overline{p}_{2}} {\sqrt{\overline{p}(1-\overline{p})\left(\frac{1}{n_{1}} + \frac{1}{n_{2}}\right)}},
    \label{eqn:z_value}
\end{equation}

\noindent
where $\overline{p}_{1}$ and $\overline{p}_{2}$ denote the average fractions (in this case, the group fraction) of the two populations being compared, with sample sizes n$_{1}$ and n$_{2}$ respectively. $\overline{p}$ denotes the so-called \textit{pooled} proportion such that;

\begin{equation}
    \overline{p} = \overline{p}_{1} \left(\frac{n_{1}}{n_{1}+n_{2}}\right) + \overline{p}_{2} \left(\frac{n_{2}}{n_{1}+n_{2}}\right).
\end{equation}

The Z-statistic (Eqn.~\ref{eqn:z_value}) is then converted to a $p$-value by taking the area under a normal distribution within which z$>$Z or z$<$Z, depending on the hypothesis being tested (see e.g. \cite{agresti98} for details). In the case of quasars in groups (Fig.~\ref{fig:group}), we obtain a $p$-value, P(z$>$Z) = 0.29, indicating that the two samples are likely drawn from the same parent population. \textit{We therefore conclude that the likelihood of a given galaxy existing in a group is not affected by the presence of a quasar, finding no statistical difference in the group environments of quasars and mass-matched inactive galaxies at low redshift.} In addition, the group multiplicity index provided by GAMA, N$_{\rm{FoF}}$, which denotes the number of galaxies associated with each group, reveals no difference in the typical group size of either population. Both the quasar and matched galaxy samples have a median group size of three, demonstrating that both populations to exist in small-to-medium sized groups.

\subsubsection{Group Centrals} \label{sec:group_centrals}

Further to identifying which galaxies exist in groups, GAMA also ranks galaxies based on their position within the group, allocated based on the distance from the group center (i.e. the central galaxy receives a rank of `1', the closest neighbor is ranked `2', the next nearest `3' and so on). Three approaches are used in GAMA to allocate these rankings and are as follows:

\begin{itemize}
    \item CoL: The center of light (CoL) is found based on the $r$-band luminosities of all galaxies associated with the group.
    \item BCG: The brightest group or cluster galaxy (BCG) is assumed to be the group center.
    \item Iter: The group center is found via an iterative process in which the galaxy at the furthest distance from the $r$-band CoL is removed. When only two group members remain, the brightest ($r$-band) galaxy is classed as the group center.
\end{itemize}

\noindent
In general, the iterative center is the preferred quantity from which to identify the group center. Based on the analysis of extensive mock catalogs, the iterative method consistently yields the closest agreement with the exact group center, returning a precise match in $\sim$90 per cent of cases, irrespective of the size of the group \citep{robotham11}. While the BCG method identifies a group center consistent with this iterative method in $\sim$95 per cent of cases for moderate group sizes (N$_{\rm{FoF}}$ $\geqslant$ 5), it is typically more sensitive to group outliers. Nevertheless, we include results from all three identifiers here for comparison. 

\begin{figure*}
\includegraphics[trim= 75 0 75 37 ,clip,width=\textwidth]{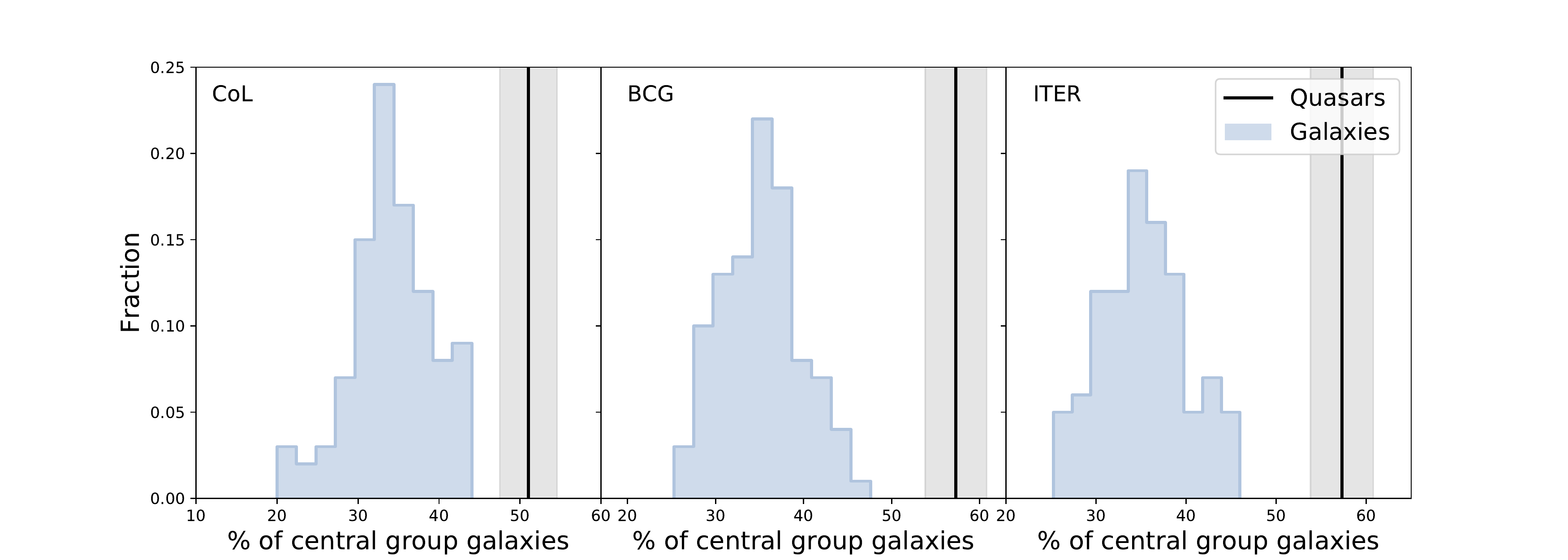}
\caption{Fraction of quasars in groups classified as the group center, based on each of the GAMA identifiers (\textit{black lines}) compared to the distribution of the fractions of group galaxies classified as the group center in each of the 100 realizations of our matched galaxy sample (\textit{blue histogram}). Shaded regions denote the uncertainty on the fraction of quasars classified as the group center, derived in Eqn.~\ref{eqn:error}.}
\label{fig:group_central}
\end{figure*}

To test the prevalence of quasars in the center of galaxy groups compared to our matched galaxy sample, we consider only the galaxies in each population found to exist in groups. According to both the 'Iter' and 'BCG' identifiers, $\sim$57 per cent of quasars in groups are the group center. Based on the 'CoL' measure, this fraction is slightly lower at $\sim$51 per cent. In any case, the probability that a quasar in a group is the center of that group, P$_{\rm{Cen}}$, exceeds 50 per cent (i.e. quasars in groups are more likely than not to be the central group galaxy). Of the group galaxies in our matched sample however, $\sim$35 per cent are classified as the group center, irrespective of the GAMA identifier chosen (see Table~\ref{tab:center}). \textit{We therefore conclude that, at low redshifts, quasars are $\sim$1.5 times more likely to exist in the center of a galaxy group than inactive galaxies of similar stellar mass} (Fig.~\ref{fig:group_central}). To test whether the observed over-representation of quasars as group centers Fig.~\ref{fig:group_central} is statistically significant, we again perform a $p$-value test (Eq.~\ref{eqn:z_value}) based on the classifications from each of the three identifiers (Iter, BCG and CoL). Indeed, we recover P(z$>$Z) $<$ 0.01 in every case, confirming the two populations are statistically distinct in terms of their group location to a confidence of $>$99 per cent. 

\begin{table}[]
    \centering
    \begin{tabular}{l|c|c}
    \hline
    Identifier &  P$_{\rm{Cen,QSO}}$ (\%) & P$_{\rm{Cen,GAL}}$ (\%) \\
    \hline
    CoL   & 51.04 $\pm$ 3.53 & 34.19 $\pm$ 4.99 \\
    BCG   & 57.29 $\pm$ 3.50 & 35.32 $\pm$ 4.56 \\
    Iter  & 57.29 $\pm$ 3.50 & 35.29 $\pm$ 4.73 \\
    \end{tabular}
    \caption{Fraction of galaxies in groups classified as the group center according to each of the three GAMA identifiers for the quasar, P$_{\rm{Cen,QSO}}$, and matched galaxy, P$_{\rm{Cen,GAL}}$, samples. Quoted uncertainties correspond to the standard error (Eq.~\ref{eqn:error}) and the standard deviation of the distribution for the quasar and matched galaxy samples respectively.}
    \label{tab:center}
\end{table}

Despite finding a clear over-representation of quasars as group centres in our sample, we note that all three identifiers in GAMA depend on the luminosity of the galaxy. Given that quasars typically appear bright across all wavelengths, often outshining their host galaxy, the luminosity dependence could result in a strong bias towards selecting quasars as the group centre. To investigate whether this is the case in our sample, we look at the 21 groups in which quasars are identified as the group centre according to the `Iter' identifier in GAMA. In 18 of these 21 groups (85.7 per cent), the central quasar is indeed shown to exhibit the brightest $r$-band magnitude in the group. Given that the the CoL is largely determined from this $r$-band magnitude, it is therefore possible that light from the quasar is strongly biasing the identifier. To address this, we subtract the quasar contribution from the $r$-band flux to determine whether their identification as the group centre is entirely dependent on the quasar emission. While modelling the true contribution of the quasar light to the overall galaxy emission lies well beyond the scope of this work, we nevertheless estimate the fractional contribution of the quasar based on the best-fit models of our CIGALE fitting (see section~\ref{sec:gama_qsos}). Of the 18 central quasars appearing as the brightest group galaxies in the $r$-band, 14 (77.8 per cent) remain the brightest after removing the contribution from the quasar. When accounting for the quasar contribution in this way and assuming the four groups for which the quasar is no longer the brightest target do not have quasars at their centre, we find a potential 12.7 per cent decrease in the number of central group quasars. This would result in 44.6 per cent of quasars in groups being identified as the group centre, rather than the 57.3 per cent given in Table ~\ref{tab:center}. We note that the corresponding increase in the fraction of our matched galaxy sample identified to be the group centre is negligible ($<$ 0.1 per cent) due to the large sample size. A $p$-vale test on these new fractions returns P(z$>$Z) = 0.0294, again confirming the two populations to be statistically distinct to a $>$97 per cent confidence. While we acknowledge that GAMA may be biased towards identifying quasars as the group center, we therefore \textbf{confirm/ stand by} the conclusion that quasars are over-represented at the centre of galaxy groups.

Our results appear broadly consistent with previous studies in which only a small fraction ($<$10 per cent) of quasars were found to exist as satellite galaxies \citep{kayo2012,richardson2012,shen2013,wang2014}. It is possible that the center of galaxy groups provides a more congenial environment when considering the triggering and fuelling of quasars, either because they experience a higher rate of mergers and galaxy interactions or because they lie at the center of cold gas flows. Although we do not find a difference in the incidence of quasars in groups compared to galaxies of the same stellar mass, the over-representation of quasars as the group center may therefore indicate that close-pair mergers, which we are not sensitive to in our local environment analysis (section.~\ref{sec:local_env}), could play an important role in the onset of nuclear activity.

\subsubsection{Radio-Detected Quasars} \label{sec:radio}

Several early studies found a difference in the environment richness of radio-loud (RLQ) and radio-quiet (RQQ) quasars, finding RLQs to be much more likely to reside in rich environments, such as group and cluster centers, compared to their radio-quiet counterparts \citep[e.g.][]{yee84, ellingson91, hintzen91, boyle93}. Likewise, more recent work by \cite{von07} suggests that the brightest group and cluster galaxies, often synonymous with the central galaxy, are more likely to be radio-loud compared to the general quasar population. Although a detailed study of the radio properties of our quasar sample lies beyond the scope of this work, we nevertheless search for evidence of an enhancement in radio detections associated with the quasars in our sample identified as the group center.

To investigate the potential link between radio emission and central group galaxies among quasars in GAMA, we positionally cross-match the 200 quasars in our sample with group information to the VLA Faint Images of the Radio Sky at Twenty-Centimeters (FIRST: \citealt{becker95}), searching for radio sources within 6.4 arcsec of our quasar targets. This search radius corresponds to the major axis of the elliptical cross-section (6.4$\times$5.4 arcsec) of the FIRST beam full width half maximum (FWHM). 31 of the 200 quasars (15.50 $\pm$ 2.56 per cent) are found to be associated with FIRST radio sources, where the uncertainty denotes the standard error on the fraction (Eq.~\ref{eqn:error}) . Although we opt to consider only the galaxies in the \textsc{GroupFindingv10} catalogue, from which the group and central fractions have been calculated (sections~\ref{sec:group_gals}-~\ref{sec:group_centrals}), we note that using the full quasar sample of 205 galaxies returns a similar fraction of radio-detected sources (15.12 $\pm$ 2.50 per cent). Similarly, we cross-match the positions of all quasars classified as the group center according to each of the GAMA identifiers. We return radio-detected fractions of 16.36 $\pm$ 4.99, 12.73 $\pm$ 4.49 and 14.29 $\pm$ 5.00 per cent for the 'Iter', 'BCG', and 'CoL' identifiers respectively. A $p$-value test (Eq. ~\ref{eqn:z_value}) returns P(z$>$Z) = 0.4368, P(z$<$Z) = 0.3050 and P(z$<$Z) based on these three identifiers, with respect to the GAMA quasars in the \textsc{GroupFindingv10} catalogue. We therefore find no statistical difference in the radio detection rates of central group quasars compared to the underlying population and thus conclude that quasars existing in the centers of groups are no more likely to be associated with bright radio sources than the general quasar population.

\subsection{Cluster Environments} \label{sec:cluster_env}

The DMU:\textsc{EnvironmentMeasuresv05} in GAMA provides several metrics of the cluster-scale galaxy environments \citep{brough13}. In this paper, we consider three such metrics: the surface density, cylinder counts and the distance to the fifth nearest neighbor. Each of these environment metrics is derived from a pseudo-volume-limited galaxy population, comprising all galaxies with an absolute magnitude, M$_{r}$ ($z_{\rm{ref}}$=0, Q=0.78) $<$ -20, where Q accounts for the redshift evolution of M$_{r}$ \citep{loveday15}. This magnitude limit corresponds to an upper limit on the redshift of $z<$0.18333, above which galaxies fulfilling this criteria in GAMA (which has an equatorial survey depth of m$_{r}<$ 19.8) become too sparse to sufficiently sample. As a result, \textsc{EnvironmentMeasuresv05} provides cluster-scale environment information only for galaxies with $z \lesssim$0.18. These environment measures are therefore provided for just 59 of the lowest redshift 205 quasars in our sample (Fig.~\ref{fig:sample_bias_env}). Despite covering a relatively narrow range of redshifts in our sample, Fig.~\ref{fig:sample_bias_env} demonstrates no bias in terms of m$_{r}$ (brightness), including all $z<$0.18 quasars from our original quasar sample. 

\begin{figure*}
	\centering
	\subfigure[]{\includegraphics[trim=0 0 0 0,clip,width=0.5\textwidth]{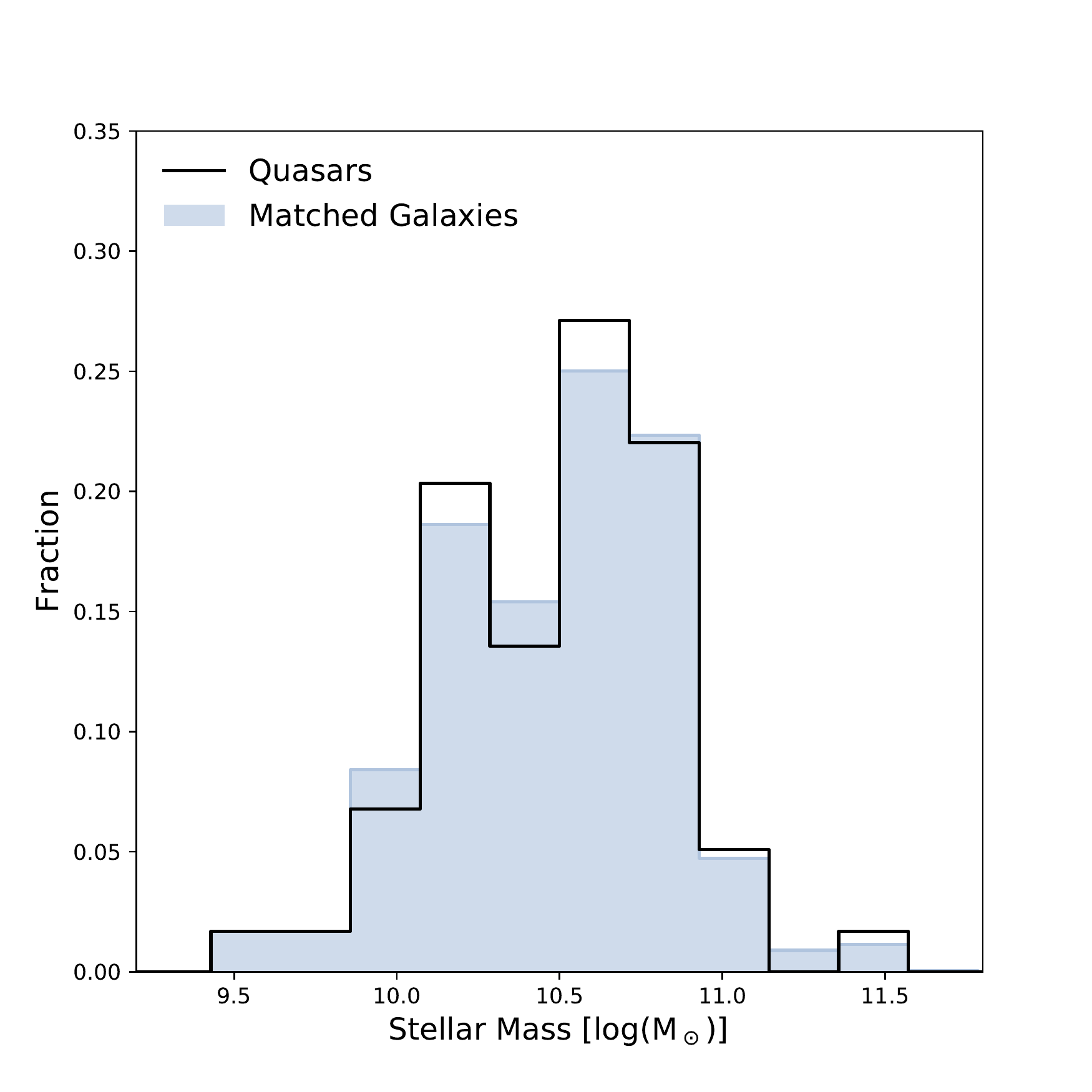}}\hfill
	\subfigure[]{\includegraphics[trim=0 0 0 0,clip,width=0.5\textwidth]{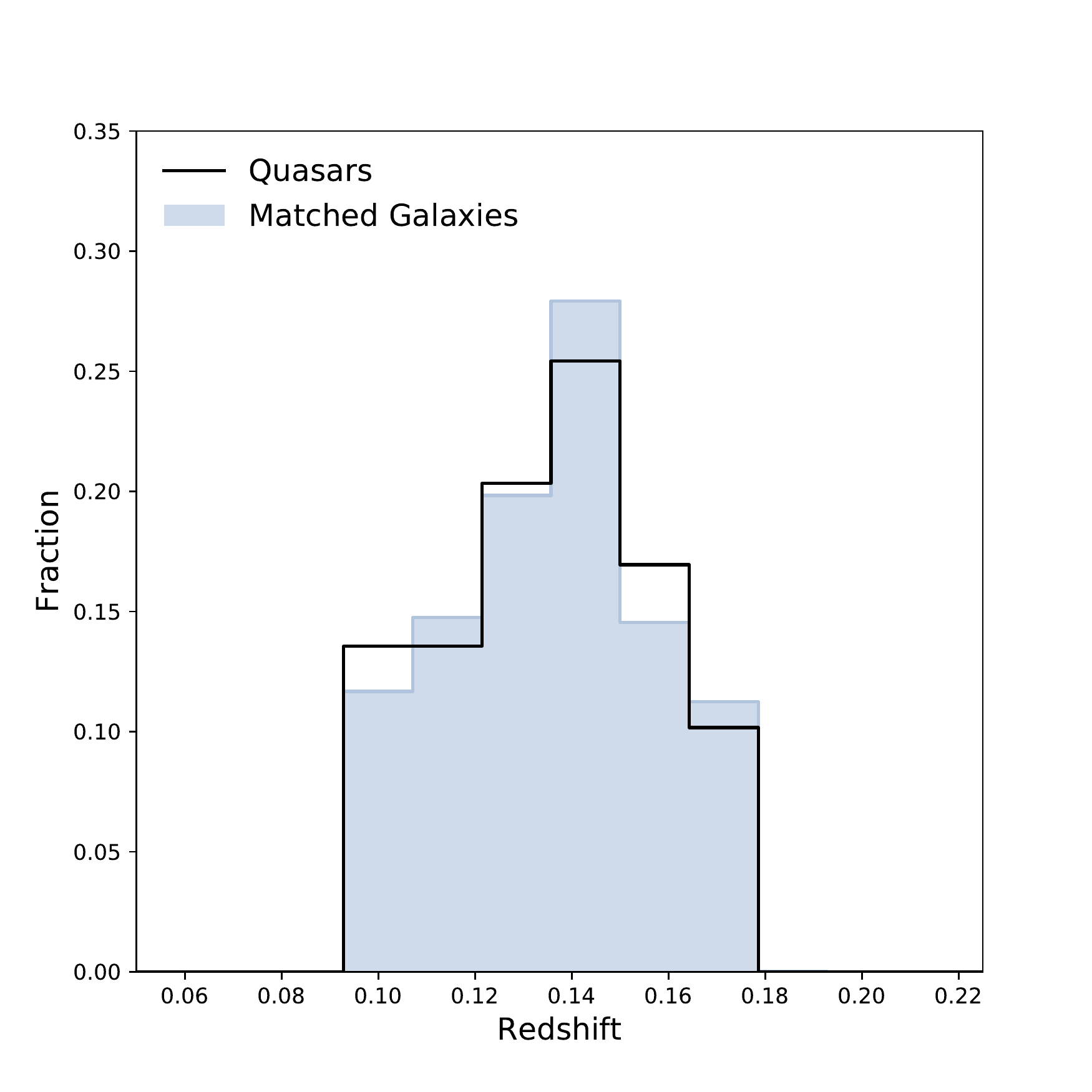}}\hfill	
\caption{Distribution of stellar masses (a) and redshifts (b) for quasars in GAMA with cluster-scale environment information (\textit{black}) and the matched galaxy sample for which this information is also available (\textit{blue}).}
\label{fig:matched_env}
\end{figure*}

To account for the significant reduction in sample size and redshift range, we re-sample 100 sets of galaxy comparison samples, each consisting of 59 GAMA galaxies matched in M$_{*}$ and redshift, following the methods outlined in section~\ref{sec:matched_gals}. The distribution of M$_{*}$ and redshift for the resulting samples (59$\times$100 galaxies) are shown in Fig.~\ref{fig:matched_env}, along with those for the 59 quasars in the redshift-limited sample.

\begin{figure}
\includegraphics[trim= 25 5 30 15 ,clip,width=.5\textwidth]{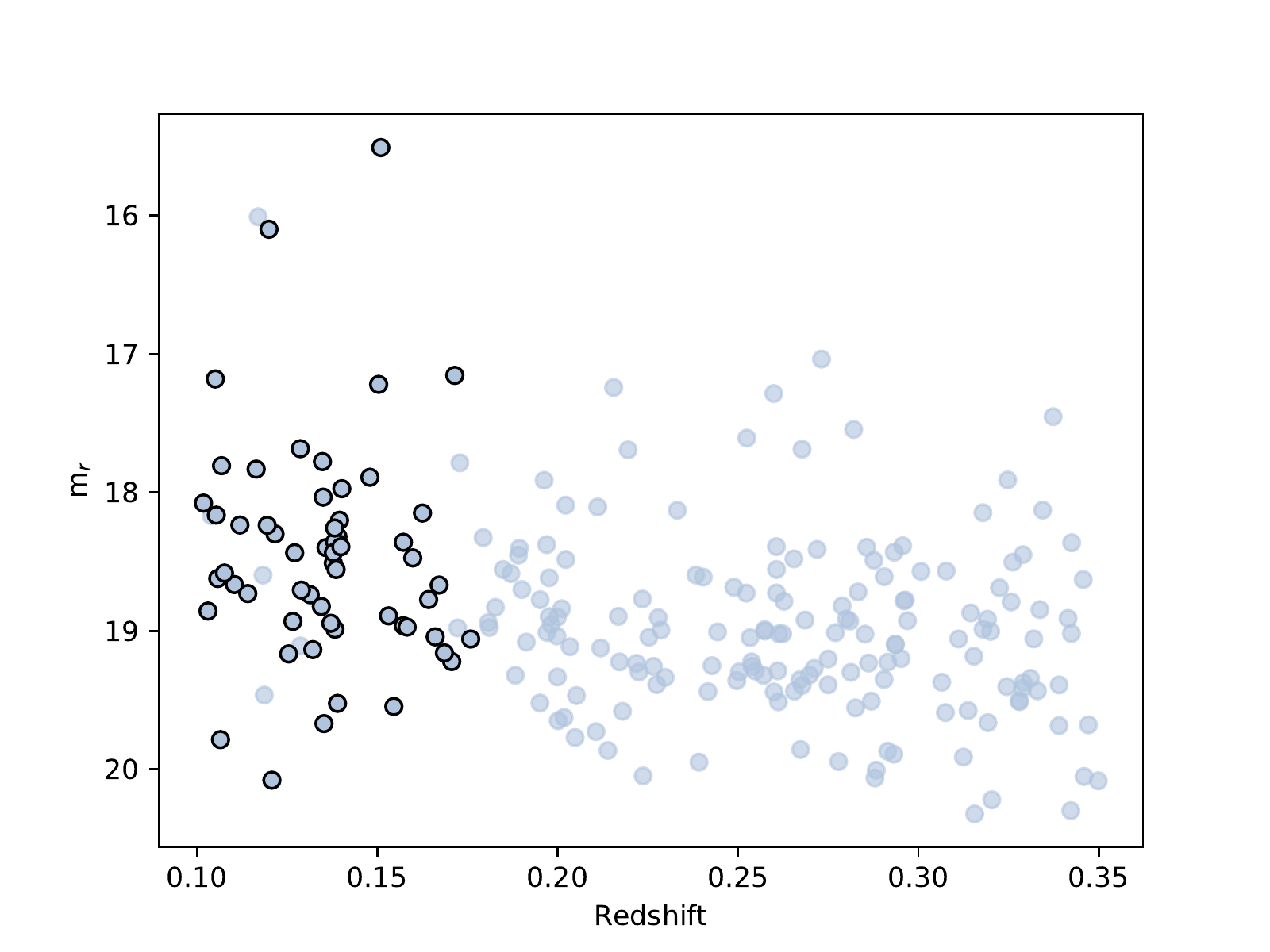}
\caption{$r$-band apparent magnitude (AB) vs. redshift for all quasar targets with cluster-scale environment information in GAMA (\textit{black circles}) compared to the full sample of quasars in LQAC-4 over the same redshift range brighter than the GAMA magnitude limit (\textit{blue dots}).}
\label{fig:sample_bias_env}
\end{figure}

\subsubsection{Surface Density} \label{sec:surface_density}

\begin{figure}
\includegraphics[trim= 110 0 100 0 ,clip,width=.5\textwidth]{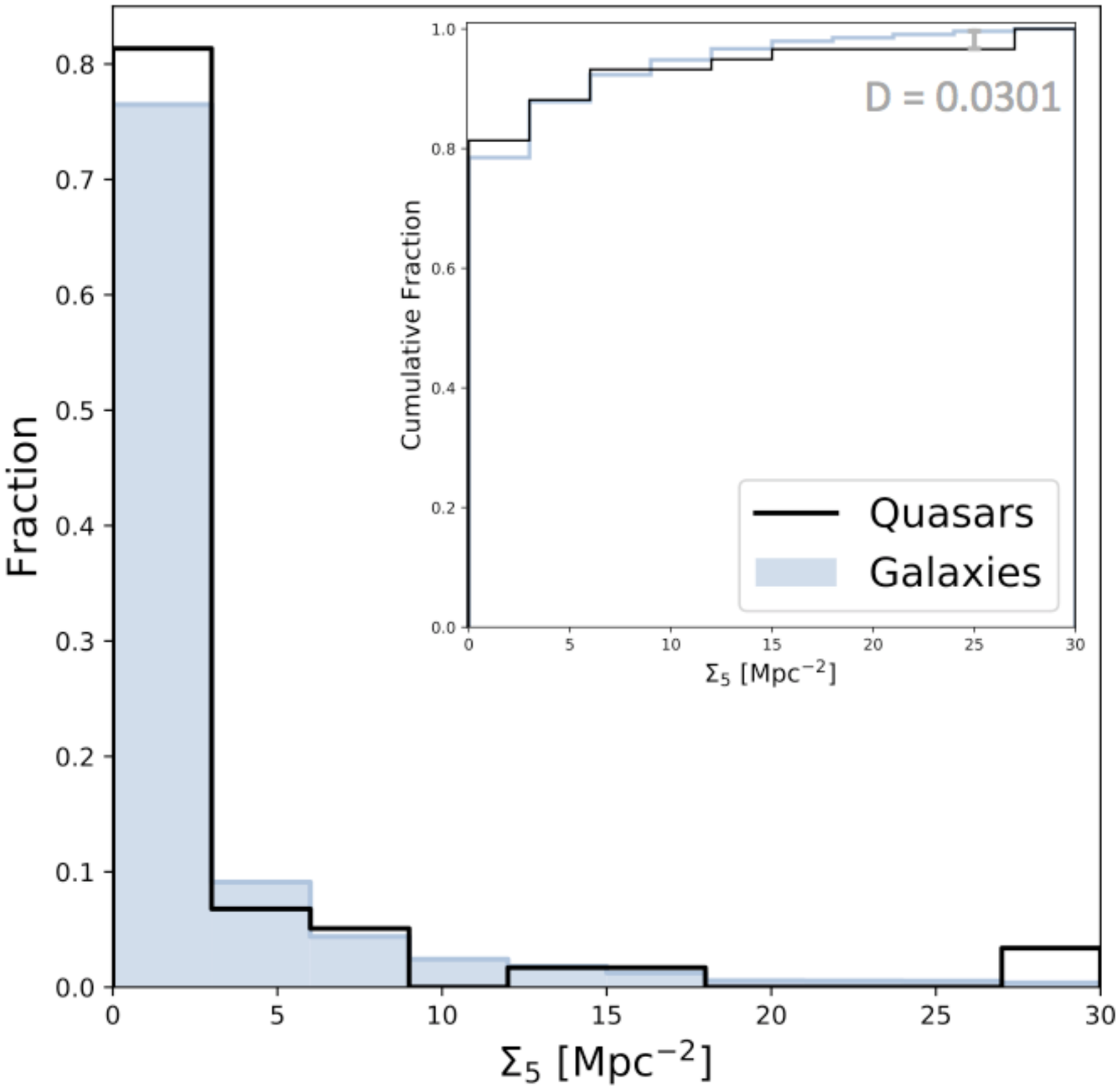}
\caption{Distribution of surface density across the GAMA quasar sample (\textit{black}) compared to that of our matched galaxy sample (\textit{blue}). Insert shows a two-sample KS test demonstrating there to be no statistical difference between the surface densities of the two populations.}
\label{fig:SD}
\end{figure}

In GAMA, the surface density, $\Sigma_{5}$, is derived from the co-moving distance (in Mpc) to the fifth nearest neighbor, $d_{5}$, within $\Delta V<$1000kms$^{-1}$, such that $\Sigma_{5}$ = 5/$\pi$$d_5$ $^2$. Full details on the derivation of $\Sigma_{5}$ are given in \citep{brough13}. Fig.~\ref{fig:SD} shows the distribution of $\Sigma_{5}$ derived in GAMA for our quasar sample and the 59$\times$100 matched galaxies. For the quasar sample we find a median average surface density, $\Sigma_{5,\rm{QSO}}$ = 0.75$^{+5.32}_{-0.17}$ Mpc$^{-2}$, compared to $\Sigma_{5,\rm{GAL}}$ = 0.48$^{+4.24}_{-0.12}$ Mpc$^{-2}$ for the matched galaxy sample. In this instance we opt to quote the median value of $\Sigma_{5,\rm{QSO}}$ with uncertainties denoting the 16th and 84th percentiles of each sample in order to account for the heavy skew of the distribution.

Although the two distributions on Fig.~\ref{fig:SD} appear similar, we nevertheless perform a two-sample Kolmogorov-Smirnov (KS) test to test their statistical consistency (see insert in Fig.~\ref{fig:SD}). The KS test is a non-parametric test used to quantify the similarity between two probability distributions \citep{massey51}, with sample sizes $n_{1}$ and $n_{2}$. In this case, we plot the cumulative frequency as a function of $\Sigma_{5}$, plotting the fraction of galaxies with a surface density $\leq$ the current value, i.e P($\leq$ $\Sigma_{5}$). The largest distance between the resulting curves, $D$, is then measured and compared to some critical value, $C(\alpha)$, of the confidence level, $\alpha$, such that  

\begin{equation}
    D < C(\alpha) \sqrt{\frac{n_{1}+n_{2}}{n_{1}n_{2}}},
\label{eqn:D}
\end{equation}

where 

\begin{equation}
    C(\alpha) = \sqrt{-ln\left(\frac{\alpha}{2}\right) \times \frac{1}{2}}.
\label{eqn:Ca}
\end{equation}

\noindent
Eqs.~\ref{eqn:D} \&~\ref{eqn:Ca} can then be rearranged to obtain the confidence level at which the two distributions differ. Typically, a result is considered statistically significant if the distributions differ with a $>$95 per cent confidence ($\alpha$ $<$ 0.05). In the case of surface densities in GAMA (Fig.~\ref{fig:SD}), we recover $\alpha$ $>$ 1, meaning \textit{we do not find any statistical difference in the surface densities of quasars and matched galaxies at low redshift.} Instead, we conclude the cluster-scale environments of both populations to be entirely consistent with one another based on their surface densities in GAMA.

\subsubsection{Cylinder Counts} \label{sec:cylinder_counts}

\begin{figure}
\includegraphics[trim= 110 0 100 0 ,clip,width=.5\textwidth]{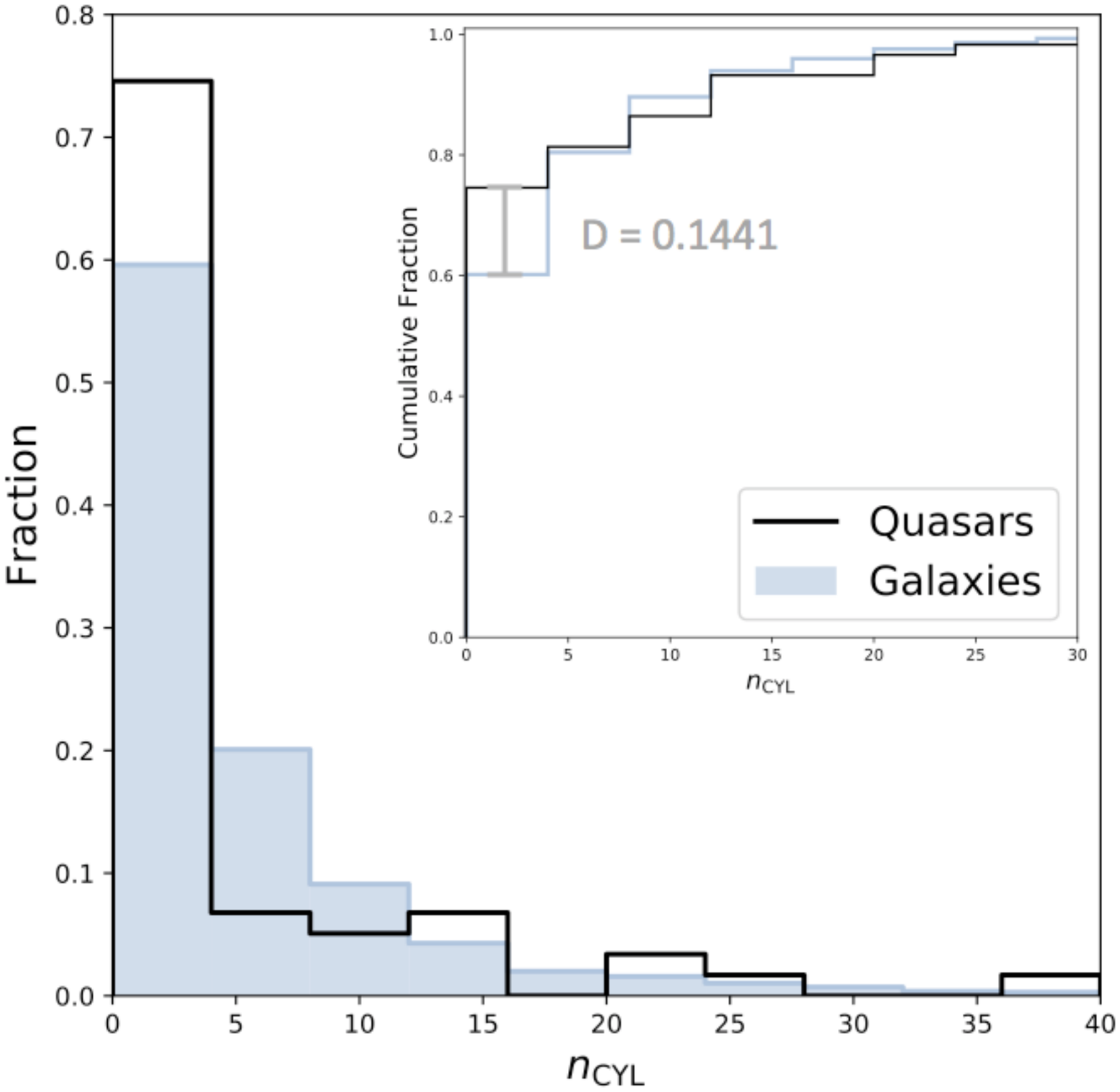}
\caption{Distribution of cylinder counts across the GAMA quasar sample (\textit{black}) compared to that of our matched galaxy sample (\textit{blue}). Insert shows a two-sample KS test demonstrating there to be no statistical difference between the cylinder counts of the two populations.}
\label{fig:Cyl}
\end{figure}

The second metric used to characterise the cluster-scale ($\sim$a few Mpc) galaxy environments is the cylinder count, $n_{\rm{CYL}}$. In GAMA, $n_{\rm{CYL}}$ denotes the number of galaxies, excluding the source itself, lying within a cylinder of radius 1 co-moving Mpc and $\Delta V<$1000kms$^{-1}$. Fig.~\ref{fig:Cyl} shows the distribution of $n_{\rm{CYL}}$ for our quasar and matched galaxy samples. Here, we derive median average values of $n_{\rm{CYL}}$ = 2.15$^{+9.33}_{-0.00}$ and $n_{\rm{CYL}}$ = 2.00$^{+10.23}_{-0.00}$ for the quasar and galaxy comparison samples respectively. Again, we opt to present the median value and 16th and 84th percentiles to account for the heavily skewed distribution. A two-sample KS test (see insert in Fig.~\ref{fig:Cyl}) returns $D$ = 0.1441. Using Eqs.~\ref{eqn:D} \&~\ref{eqn:Ca}, we calculate the the significance at which the two distributions differ, returning $\alpha$ = 0.1768. We therefore cannot rule out the similarity of the two populations with more than an 82.32 per cent confidence, which is not statistically significant. Thus, we again find the cluster-scale environments of quasars to be statistically consistent with the matched galaxy sample.

\subsubsection{Distance to Fifth Nearest Neighbor} \label{sec:fifth_neighbor}

\begin{figure}
\includegraphics[trim= 110 0 100 0 ,clip,width=.5\textwidth]{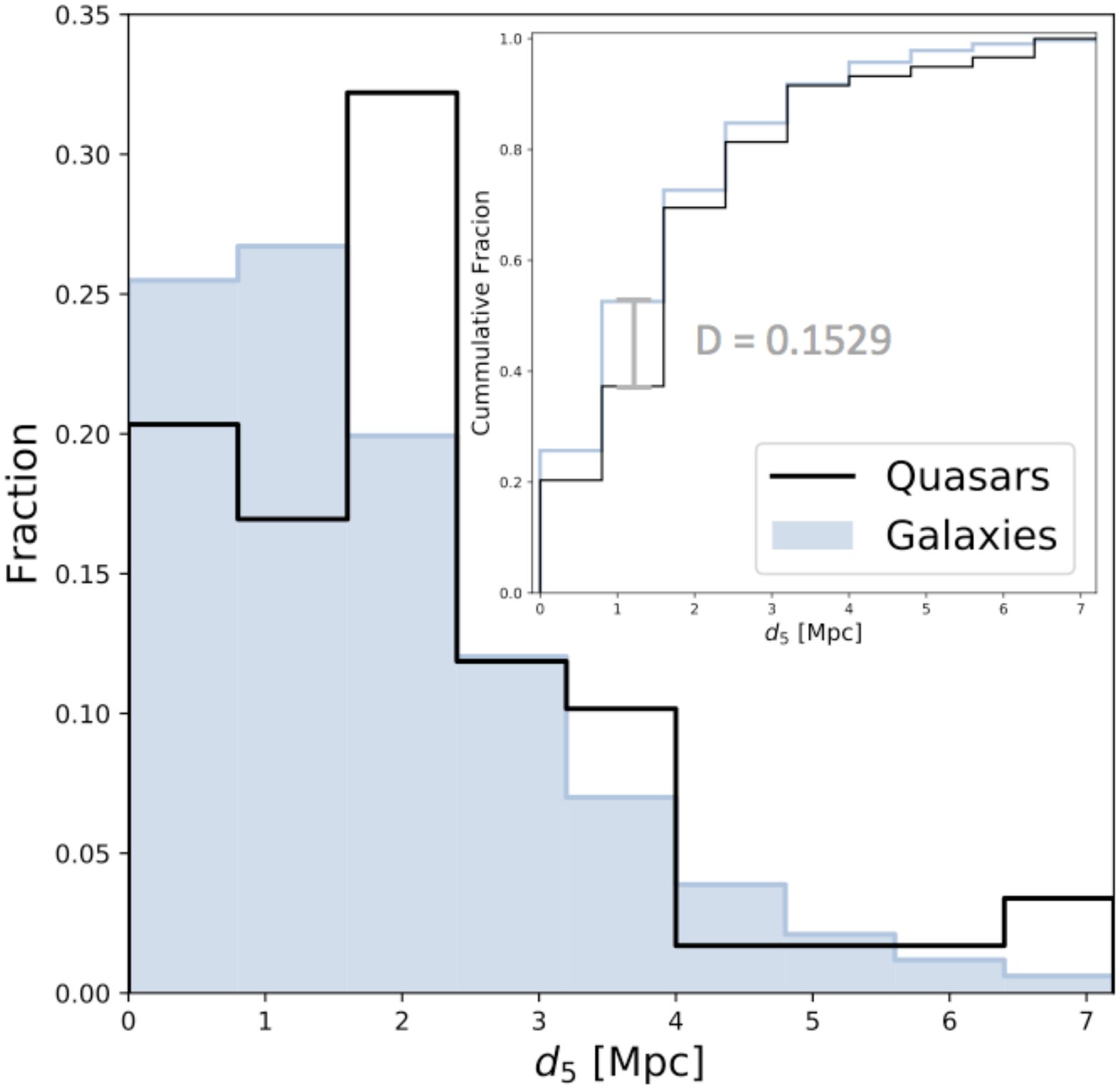}
\caption{Distribution of distances to the fifth nearest neighbor for the GAMA quasar sample (\textit{black}) compared to that of our matched galaxy sample (\textit{blue}). Insert shows a two-sample KS test indicating no statistical difference between the two distributions.}
\label{fig:5NN}
\end{figure}

The third and final metric we present here is the distance to the fifth nearest neighbor, $d_{5}$, measured in units of Mpc. The distribution of $d_5$ across our quasar sample is plotted alongside that of the matched galaxy sample in Fig.~\ref{fig:5NN}. On average, we find quasars to have $d_{5,\rm{QSO}}$ = 1.52$^{+3.20}_{-0.57}$ Mpc, compared to $d_{5,\rm{GAL}}$ = 1.86$^{+3.53}_{-0.62}$ Mpc for the matched galaxy sample. Once again, the median and the 16th and 84th percentiles have been selected to denote the average and the associated uncertainties, accounting for the heavy skew of the distributions. Although the average values of $d_5$ appear similar (consistent within the 1$\sigma$ uncertainties), we note that there exist comparatively few quasars with their fifth-nearest neighbor at the lowest separations ($\lesssim$1.5 Mpc). 

To test whether the difference observed here is indeed significant, we again perform a two-sample KS test (Fig.~\ref{fig:5NN} insert). Here, we derive a value of $D$ = 0.1529, corresponding to $\alpha$ = 0.1303 (Eqs.~\ref{eqn:D} \& ~\ref{eqn:Ca}). We therefore rule out the two populations being drawn from the same underlying distribution with a confidence of 86.97 per cent. Although this once again fails our outlined requirements for statistical significance ($>$95 per cent), we note that the difference here is stronger than in the case of either $\Sigma_{5}$ or $n_{\rm{CYL}}$. While we conclude $d_{5,\rm{GAL}}$ of the two populations to be consistent with one another, this result may therefore tentatively suggest a preference for quasars existing in intermediate-density cluster environments, with comparatively few quasars existing in the densest cluster environments ($d_{5,\rm{GAL}}$ $<$1.5 Mpc).

\subsection{Large Scale Structure}

In addition to providing information on the local, group and cluster environments of galaxies, GAMA classifies galaxies into one of four large-scale ($>$10 Mpc) structures: voids, sheets, filaments and knots, based on the galaxy’s so-called deformation (or tidal) tensor, $T_{ij}$ \citep{eardley15}. Broadly speaking, the number of positive eigenvalues for $T_{ij}$ indicate whether the structure is collapsing in zero (void), one (sheet), two (filament) or three (knot) dimensions. In GAMA, $T_{ij}$ is computed with both a 4 Mpch$^{-1}$ and a 10 Mpch$^{-1}$ smoothing, which we shall hereafter refer to as the GeoS4 and GeoS10 classifiers respectively.  

\begin{figure}
\includegraphics[trim= 25 5 30 15 ,clip,width=.5\textwidth]{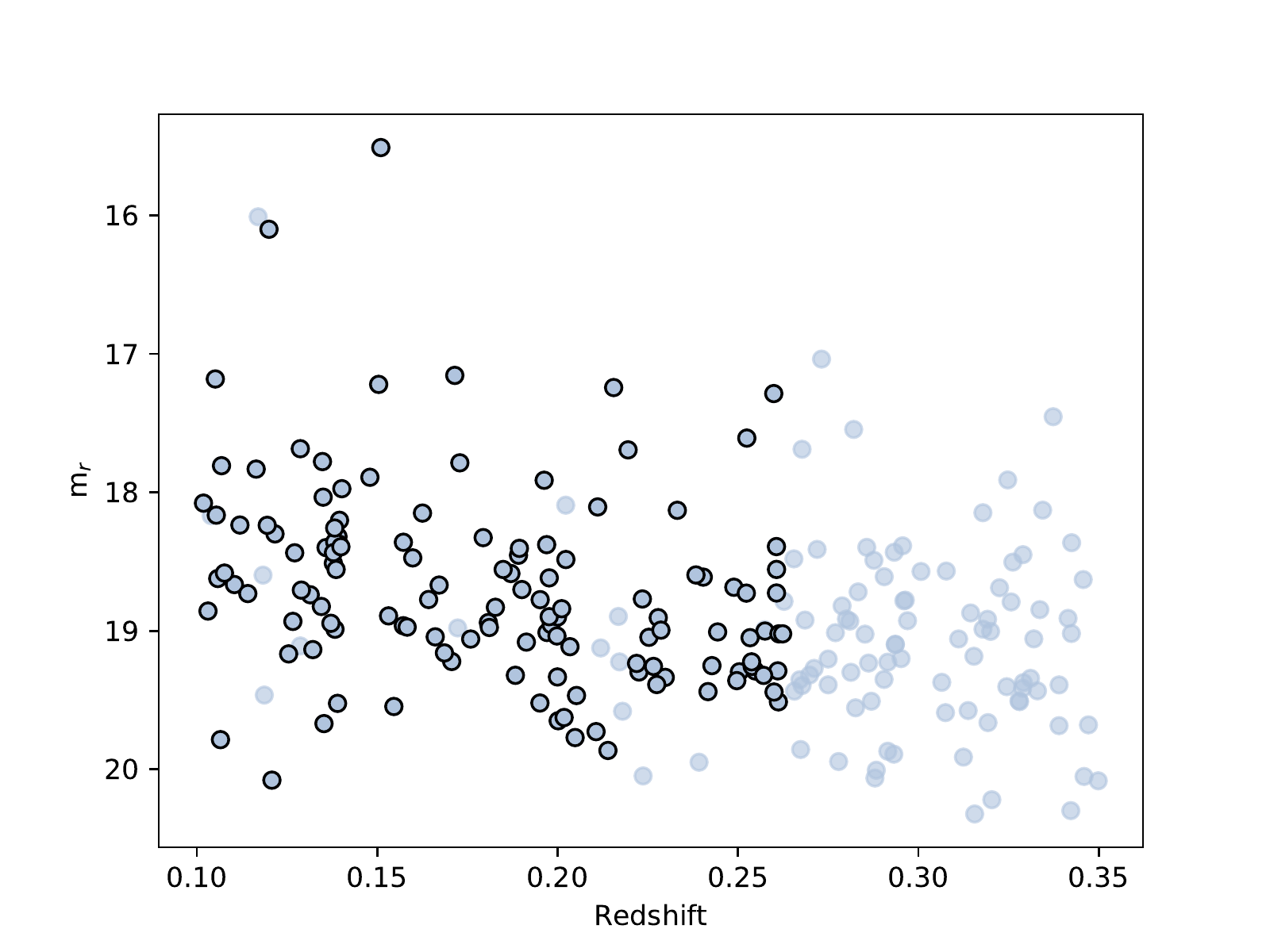}
\caption{$r$-band apparent magnitude (AB) vs. redshift for all quasar targets with LSS information in GAMA (\textit{black circles}) compared to the full sample of quasars in LQAC-4 over the same redshift range brighter than the GAMA magnitude limit (\textit{blue dots}).}
\label{fig:sample_bias_lss}
\end{figure}

\begin{figure*}
	\centering
	\subfigure[]{\includegraphics[trim=0 0 0 0,clip,width=0.5\textwidth]{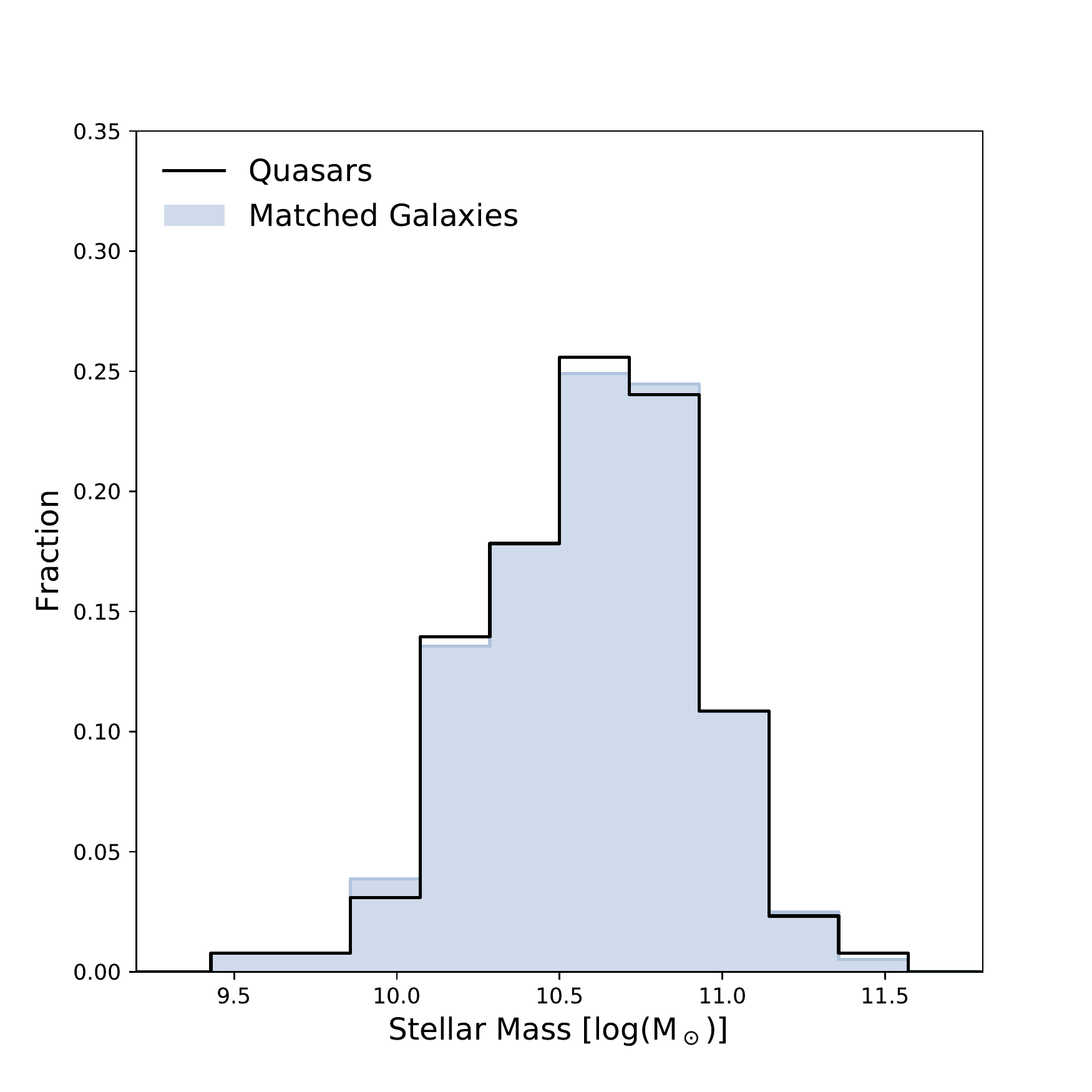}}\hfill
	\subfigure[]{\includegraphics[trim=0 0 0 0,clip,width=0.5\textwidth]{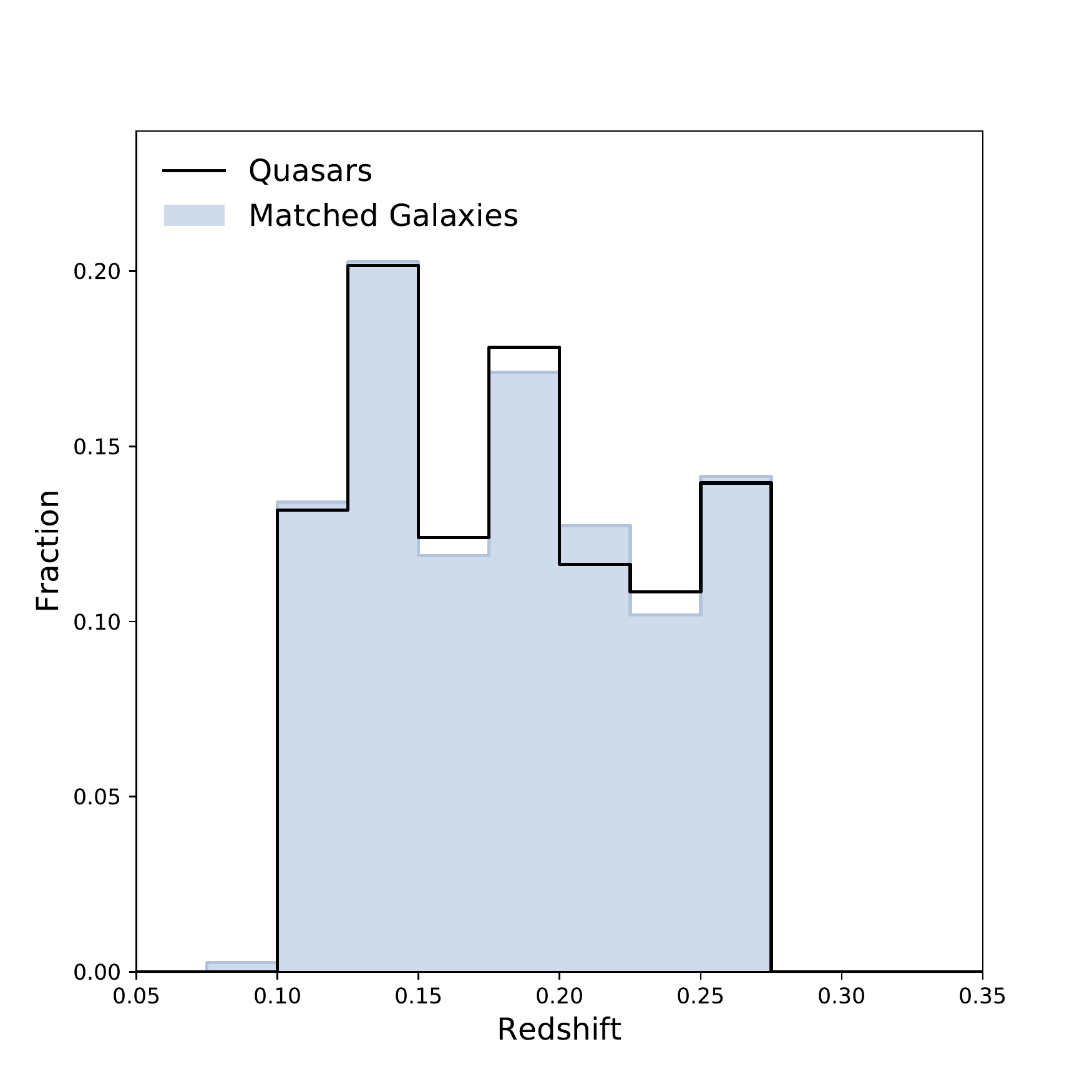}}\hfill	
\caption{Distribution of stellar masses (a) and redshifts (b) for quasars in GAMA with LSS information (\textit{black}) and the matched galaxy sample for which this information is also available (\textit{blue}).}
\label{fig:matched_lss}
\end{figure*}

As for the cluster environments (section~\ref{sec:cluster_env}), the large-scale structure can only be derived for galaxies to L$*$. As we move to higher redshifts, galaxies appear fainter, meaning that despite the large volume, relatively few galaxies are detected in GAMA at the highest redshifts in our sample ($z\gtrsim$0.26). As a result, the galaxy sampling at these redshifts is too sparse to derive information about their large-scale structure. Information on the large-scale structure is therefore available in GAMA for 129 of the 205 quasars in our sample, with a clear bias towards lower-redshift systems (Fig.~\ref{fig:sample_bias_lss}), although we note that this effect is far less severe than for the cluster environments. To account for this reduction in sample size and redshift range, we again re-sample sets of matched galaxies, identifying 100 realisations (100$\times$129) of a galaxy sample matched in stellar mass and redshift to the reduced quasar sample (Fig.~\ref{fig:matched_lss}). 

\begin{figure*}
\includegraphics[trim= 115 0 120 37 ,clip,width=\textwidth]{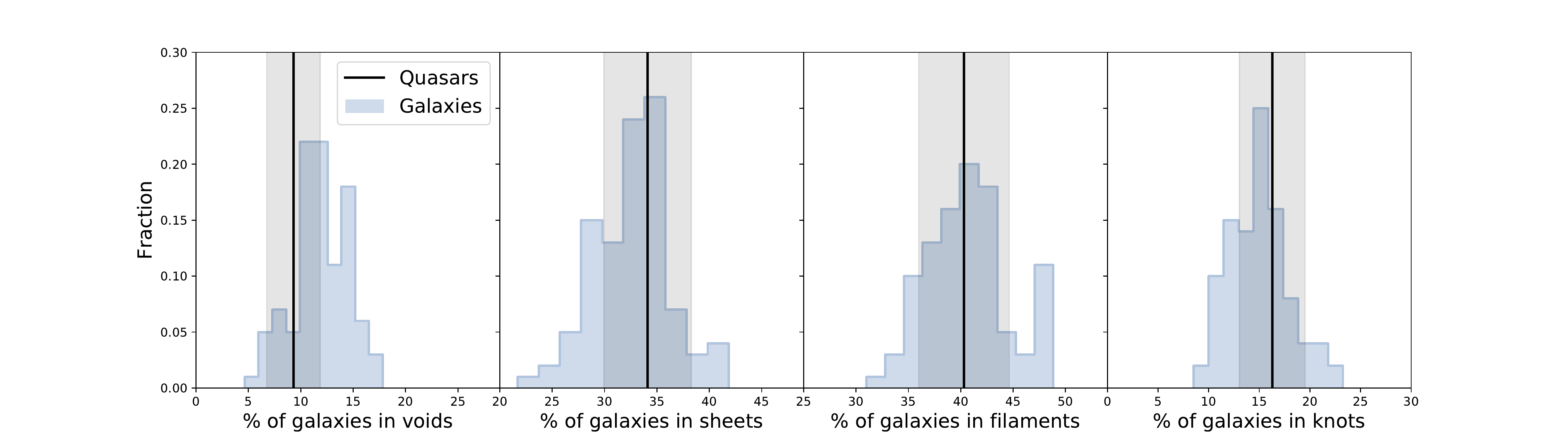}
\caption{Fraction of quasars in GAMA found to exist in each LSS environment (void, sheet, filament and knot) based on the GeoS4 classifier (\textit{black}) compared to the distribution of the fractions across the 100 realizations of our matched galaxy sample (\textit{blue histogram}). Shaded regions denote the uncertainty on the fraction of quasars in each environment, derived in Eqn.~\ref{eqn:error}.}
\label{fig:lss_GeoS4}
\end{figure*}

Based on the GeoS4 large-scale structure information in GAMA, we calculate the fraction of our quasar sample found to exist in voids, sheets, filaments and knots. These fractions are given in Tab.~\ref{tab:lss_s4}, along with the corresponding fractions for the matched galaxy sample. The fraction of quasars in each environment is also plotted in Fig.~\ref{fig:lss_GeoS4} alongside the corresponding distributions derived from each of the 100 galaxy samples. In general, both populations appear to prefer intermediate-density sheets and filaments to either very high density knots or low density voids. The biggest difference between the two populations is seen in the low-density voids, where quasars appear less likely to exist in voids than the matched galaxy sample, potentially indicating that quasars reside in slightly denser environments than galaxies of the same stellar mass. We note however, that even here the difference is small ($<$3 per cent). A $p$-value test confirms that the difference between the large-scale environments of each population are statistically insignificant in every case (see Tab.~\ref{tab:lss_s4}). 

\begin{table}[]
    \centering
    \begin{tabular}{l|c|c|c}
    \hline
    Environment &  P$_{\rm{QSO,S4}}$ (\%) & P$_{\rm{GAL,S4}}$ (\%) & P$_{\rm{S4}}$(z$<$/$>$Z)\\
    \hline
    Void     &  9.30 $\pm$ 2.56 & 11.92 $\pm$ 2.59 & 0.18 \\
    Sheet    & 34.11 $\pm$ 4.17 & 32.55 $\pm$ 3.77 & 0.35 \\
    Filament & 40.31 $\pm$ 4.32 & 40.62 $\pm$ 3.94 & 0.47 \\
    Knot     & 16.28 $\pm$ 3.25 & 14.91 $\pm$ 3.06 & 0.33 \\
    \end{tabular}
        \caption{Fraction of galaxies existing in each LSS environment based on the GeoS4 classifier in GAMA for the quasar, P$_{\rm{QSO,S4}}$, and matched galaxy, P$_{\rm{GAL,S4}}$, samples, along with the corresponding $p$-value for each environment. Quoted uncertainties denote the standard error (Eq.~\ref{eqn:error}) and the standard deviation of the distribution for the quasar and matched galaxy samples respectively.}
    \label{tab:lss_s4}
\end{table}

\begin{figure*}
\includegraphics[trim= 115 0 120 37 ,clip,width=\textwidth]{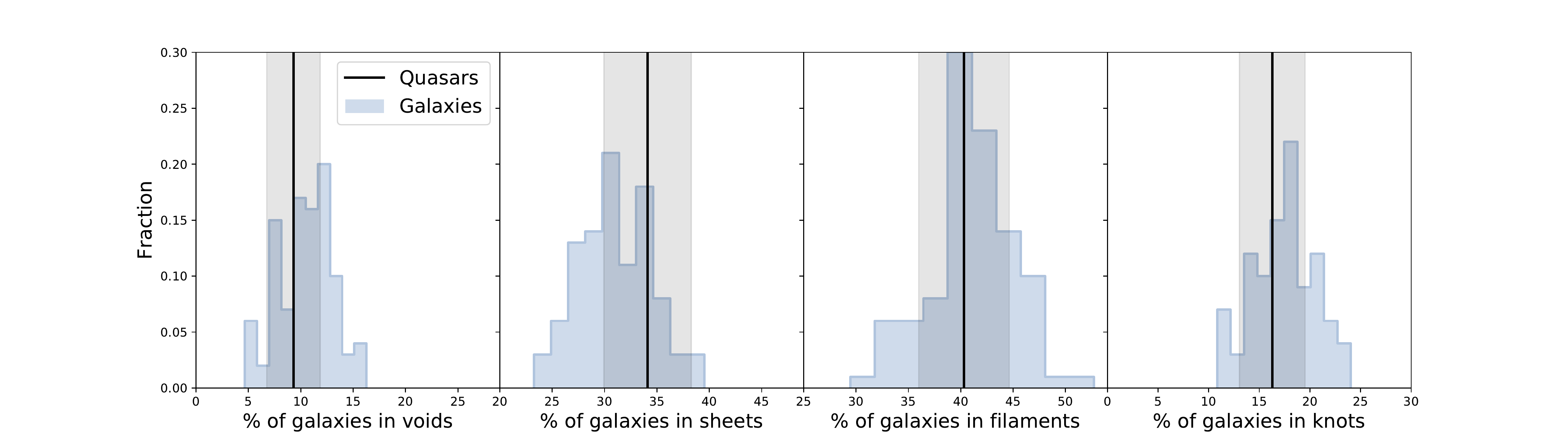}
\caption{Fraction of quasars in GAMA found to exist in each LSS environment (void, sheet, filament and knot) based on the GeoS10 classifier (\textit{black}) compared to the distribution of the fractions across the 100 realizations of our matched galaxy sample (\textit{blue histogram}). Shaded regions denote the uncertainty on the fraction of quasars in each environment, derived in Eqn.~\ref{eqn:error}.}
\label{fig:lss_GeoS10}
\end{figure*}

Perhaps unsurprisingly, the GeoS10 classifier returns similar results. Again, both quasars and the matched galaxies are shown to preferentially reside in sheets and filaments, with relatively few galaxies existing in knots or voids in either sample (Tab.~\ref{tab:lss_s10}, Fig.~\ref{fig:lss_GeoS10}). This result is consistent with several previous studies in which quasars have been shown to avoid both very over- and under-dense environments, typically residing in intermediate-density regions in terms of their large scale structure \citep[e.g.][]{miller03,villforth12}. Furthermore, as was the case for the GeoS4 classifications, a $p$-value test demonstrates the two populations to be entirely consistent with one another in terms of their GeoS10 large-scale structure, finding no statistical difference. \textit{We therefore conclude that quasars at 0.10$<$z$<$0.26 do not trace the large-scale structure of the universe any more than `typical' galaxies of the same stellar mass.} 

\begin{table}[]
    \centering
    \begin{tabular}{l|c|c|c}
    \hline
    Environment &  P$_{\rm{QSO,S10}}$ (\%) & P$_{\rm{GAL,S10}}$ (\%) & P$_{\rm{S10}}$(z$<$/$>$Z)\\
    \hline
    Void     &  9.30 $\pm$ 2.56 & 10.28 $\pm$ 2.55 & 0.36 \\
    Sheet    & 34.11 $\pm$ 4.17 & 30.99 $\pm$ 3.27 & 0.22 \\
    Filament & 40.31 $\pm$ 4.32 & 41.22 $\pm$ 4.03 & 0.42 \\
    Knot     & 16.28 $\pm$ 3.25 & 17.51 $\pm$ 3.08 & 0.36 \\
    \end{tabular}
       \caption{Fraction of galaxies existing in each LSS environment based on the GeoS10 classifier in GAMA for the quasar, P$_{\rm{QSO,S10}}$, and matched galaxy, P$_{\rm{GAL,S10}}$, samples, along with the corresponding $p$-value for each environment. Quoted uncertainties denote the standard error (Eq.~\ref{eqn:error}) and the standard deviation of the distribution for the quasar and matched galaxy samples respectively.}
    \label{tab:lss_s10}
\end{table}

\subsubsection{Fisher's Exact Test}

In addition to the above $p$-value analysis, we perform Fisher's exact test \citep{fisher58}. This test is used to quantify the significance of association, or \textit{contingency}, between two sets of categorical data. In the case of this work, we consider the quasar and matched galaxy samples as the two data sets. Unlike other statistical tests, e.g. chi-squared, which produce only an approximation, the Fisher test provides an exact result, meaning it can be applied to much smaller data sets, such as our quasar sample. Here, we perform Fisher's exact test on 2$\times$4 contingency tables containing the number of quasars, N$_{\rm{QSO}}$, and matched galaxies, N$_{\rm{GAL}}$, classified into each of the four large-scale structure environments in GAMA. For the quasars, N$_{\rm{QSO}}$ corresponds simply to the number of targets in our sample (129 quasars) that are classified as existing in voids, sheets, filaments or knots. In the case of the matched galaxy sample, N$_{\rm{GAL}}$ instead denotes the median number of galaxies existing in each environment for a given set of 129 galaxies across each of the 100 sample realisations. The contingency tables for the GAMA GeoS4 and GeoS10 classifiers are given in Tab.~\ref{tab:contingency1} and Tab.~\ref{tab:contingency2} respectively. 

\begin{table}[]
    \centering
    \begin{tabular}{l|c|c|c}
    \hline
    Environment &  N$_{\rm{QSO,S4}}$ & N$_{\rm{GAL,S4}}$ & Row Total\\
    \hline
    Void     & 12 & 16 & 28 \\
    Sheet    & 44 & 42 & 86 \\
    Filament & 52 & 52 & 104 \\
    Knot     & 21 & 19 & 40 \\
    \hline
    Column Total & 129 & 129 & 258 \\
    \end{tabular}
       \caption{Contingency table based on the GeoS4 classifier used in the Fisher's exact test.}
    \label{tab:contingency1}
\end{table}

\begin{table}[]
    \centering
    \begin{tabular}{l|c|c|c}
    \hline
    Environment &  N$_{\rm{QSO,S10}}$ & N$_{\rm{GAL,S10}}$ & Row Total\\
    \hline
    Void     & 12 & 13 & 25 \\
    Sheet    & 44 & 40 & 84 \\
    Filament & 52 & 53 & 105 \\
    Knot     & 21 & 23 & 44 \\
    \hline
    Column Total & 129 & 129 & 258 \\
    \end{tabular}
       \caption{Contingency table based on the GeoS10 classifier used in the Fisher's exact test.}
    \label{tab:contingency2}
\end{table}

Based on the contingency tables (Tab.~\ref{tab:contingency1} and Tab.~\ref{tab:contingency2}), we calculate the point probability using Simple Interactive Statistical Analysis (SISA). For the case of the GeoS4 classifier (Tab.~\ref{tab:contingency1}), we recover a two-sided probability of P(O$\geq$E$|$O$\leq$E) $=$ 0.8745, where O and E denote the observed and expected probabilities respectively. This result shows no statistical difference in the large-scale environments of quasars compared to the matched galaxy sample, indicating instead that the two data sets are likely drawn from the same underlying population. Likewise, we find no statistical difference in the large scale environments of quasars compared to the matched galaxy sample in terms of their GeoS10 classifications in GAMA (Tab.~\ref{tab:contingency2}), for which the test returns P(O$\geq$E$|$O$\leq$E) $=$ 0.95769.

\section{Conclusions} \label{sec:conclusions}

The role of galaxy environment in triggering quasar activity remains a key question in building a cohesive picture of galaxy-quasar coevolution. Throughout this work, we have explored the connection between quasar activity and galaxy environment through a direct comparison of quasars in GAMA with a set of galaxies matched in both stellar mass and redshift to the quasar sample. Our key conclusions are as follows;

i) On scales $<$100 kpc, we find no difference in the environments of the quasar and matched galaxy samples, which return an average galaxy neighbor count of 0.22$\pm$0.03 and 0.24$\pm$0.04 sources respectively within $<$100 kpc and $\Delta V<$1000 kms$^{-1}$. Although we cannot rule out close pairs within $\sim$10 kpc due to the convolution limit of SDSS, we find no evidence for an overdensity in quasar environments on scale $<$100 kpc. By extension, we suggest that major gas-rich mergers, often associated with such overdensities on these small scales, are unlikely to be the dominant triggering mechanism for the quasars in our sample. Rather, this result suggests that secular processes are likely to play a much more important role in triggering quasar activity at low redshift.

ii) When comparing the group-scale (sub Mpc) environments of our quasar and matched galaxy samples, we again find no statistical differences between the two populations. According to the group identifier in GAMA, 48.00 $\pm$ 3.53 per cent of quasars reside in groups, compared to 46.07 $\pm$ 3.57 per cent of matched galaxies. A $p$-value test returns P(z$>$Z) = 0.29, indicating the two samples are likely drawn from the same parent population in terms of their group environment. On average, both populations appear to reside in small-to-moderate groups, with a median group size of three in both cases. The lack of connection between group membership and quasar activity supports the idea that galaxy interactions are not required to trigger quasars at 0.1$< z <$0.35.

iii) Although GAMA quasars do not preferentially reside in groups compared to our matched galaxy sample, they are $\sim$1.5 times more likely to be identified as the group center. Based on the three identifiers in GAMA (Iter, BCG and CoL), $>$50 per cent of quasars in groups are classified as the central galaxy, compared to $\sim$35 per cent of the matched galaxy sample. This over-representation of quasars as the group center is found to be statistically significant, returning P(z$>$Z) $<$ 0.01. Our results therefore suggest that the center of galaxy groups may provide a more congenial environment in terms of triggering and fuelling quasar activity, potentially due to the presence of cold gas flows or higher rates of galaxy interactions. Given that we detect no overdensity in the $<$100 kpc environments of quasars, we suggest that perhaps close-pair mergers may play a role triggering quasar activity. Despite claims that quasars residing in the center of galaxy groups are more likely to be radio-loud, we find no difference in the fraction of central quasars associated with radio emission in FIRST compared to our full quasar sample.

iv) On galaxy cluster scales of $\sim$a few Mpc, we find no statistical difference in either the surface density, $\Sigma_{5}$, or cylinder counts, $n_{\rm{CYL}}$, of quasars compared to the matched galaxy sample. This lack of distinction between the two samples in terms of their cluster-scale environments implies that nuclear activity is, at best, weakly correlated with environment on scales of a few Mpc. This result also indicates that quasars do not preferentially trace galaxy clusters compared to other galaxies matched in stellar mass. Despite the similarity in the cluster membership rate of each population, the GAMA fifth nearest neighbor distance indicates a slight preference for quasars existing in intermediate cluster environments, with comparatively few quasars having five neighbors lying at $<$1.5 Mpc.

v) Finally, beyond 10 Mpc we find no difference in the large-scale structure environments of our quasar sample compared to those of the matched galaxies. Both populations exist predominantly in intermediate-density sheets and filaments, with relatively few galaxies from either sample found in very low- or high-density voids or knots. Our results therefore suggest that low redshift quasars do not preferentially trace the large-scale structure of the universe compared to galaxies of similar stellar mass, implying that nuclear activity and galaxy environment are independent on scales $>$10 Mpc.

Overall, we find the environments of quasars in GAMA to be consistent with the general galaxy population when matched in stellar mass and redshift. This holds true across their local, group and cluster environments, as well as the large scale structures in which they reside. Unlike at higher redshifts, where major gas-rich mergers are thought to be required to trigger quasar activity \citep[e.g.][]{hopkins07,sanders88}, the lack of an environmental enhancement on local scales around quasars contradicts this merger-driven quasar paradigm at low redshift. Although we cannot explicitly rule out close pair mergers within $\lesssim$10 kpc, the strong similarities between quasar environments and those of our matched galaxy sample out to 10 Mpc, suggests that mergers are not the dominant trigger of quasar activity in our sample. Instead we suggest that secular processes, such as bar instabilities and stochastic gas accretion, may play a much larger role in triggering the quasar phenomenon at 0.1$< z <$0.35. Furthermore, this consistency between the environments of both populations supports an evolutionary picture of quasars, in which they are not an intrinsically distinct class of objects, rather a phase in the lifetime of massive galaxies. 


\section*{Acknowledgements}

CFW and JK acknowledge financial support from the Academy of Finland, grant 311438. SB acknowledges funding support from the Australian Research Council through a Future Fellowship (FT140101166). GAMA is a joint European-Australasian project based around a spectroscopic campaign using the Anglo-Australian Telescope. The GAMA input catalogue is based on data taken from the Sloan Digital Sky Survey and the UKIRT Infrared Deep Sky Survey. Complementary imaging of the GAMA regions is being obtained by a number of independent survey programmes including GALEX MIS, VST KiDS, VISTA VIKING, WISE, Herschel-ATLAS, GMRT and ASKAP providing UV to radio coverage. GAMA is funded by the STFC (UK), the ARC (Australia), the AAO, and the participating institutions. The GAMA website is http://www.gama-survey.org/.

\section*{Data Availability}

Except where otherwise stated, the data underlying this article are publicly available via the online GAMA database at \url{http://www.gama-survey.org}. In section~\ref{sec:local_env} we make specific use of version 7 of the GAMA input catalogue from this database. The radio data for the sample used in Section~\ref{sec:radio} are publicly available from the VLA FIRST archive at \url{http://sundog.stsci.edu}.



\bibliography{main}{}

\begin{thebibliography}{}
\expandafter\ifx\csname natexlab\endcsname\relax\def\natexlab#1{#1}\fi
\providecommand{\url}[1]{\href{#1}{#1}}
\providecommand{\dodoi}[1]{doi:~\href{http://doi.org/#1}{\nolinkurl{#1}}}
\providecommand{\doeprint}[1]{\href{http://ascl.net/#1}{\nolinkurl{http://ascl.net/#1}}}
\providecommand{\doarXiv}[1]{\href{https://arxiv.org/abs/#1}{\nolinkurl{https://arxiv.org/abs/#1}}}

\bibitem[{Agresti \& Coull(1998)}]{agresti98}
Agresti, A., \& Coull, B.~A. 1998, The American Statistician, 52, 119

\bibitem[{Bahcall {et~al.}(1969)Bahcall, Schmidt, \& Gunn}]{bahcall69}
Bahcall, J.~N., Schmidt, M., \& Gunn, J.~E. 1969, The Astrophysical Journal,
  157, L77

\bibitem[{Bahcall \& Chokshi(1991)}]{bahcall91}
Bahcall, N.~A., \& Chokshi, A. 1991, The Astrophysical Journal, 380, L9

\bibitem[{Baldry {et~al.}(2018)Baldry, Liske, Brown, Robotham, Driver, Dunne,
  Alpaslan, Brough, Cluver, Eardley, {et~al.}}]{baldry18}
Baldry, I., Liske, J., Brown, M., {et~al.} 2018, Monthly Notices of the Royal
  Astronomical Society, 474, 3875

\bibitem[{Baldry {et~al.}(2010)Baldry, Robotham, Hill, Driver, Liske, Norberg,
  Bamford, Hopkins, Loveday, Peacock, {et~al.}}]{baldry10}
Baldry, I.~K., Robotham, A.~S., Hill, D.~T., {et~al.} 2010, Monthly Notices of
  the Royal Astronomical Society, 404, 86

\bibitem[{Balogh {et~al.}(2004)Balogh, Baldry, Nichol, Miller, Bower, \&
  Glazebrook}]{balogh04}
Balogh, M.~L., Baldry, I.~K., Nichol, R., {et~al.} 2004, The Astrophysical
  Journal Letters, 615, L101

\bibitem[{Barnes \& Hernquist(1992)}]{barnes92}
Barnes, J.~E., \& Hernquist, L. 1992, Annual review of astronomy and
  astrophysics, 30, 705

\bibitem[{Becker {et~al.}(1995)Becker, White, \& Helfand}]{becker95}
Becker, R.~H., White, R.~L., \& Helfand, D.~J. 1995, The Astrophysical Journal,
  450, 559

\bibitem[{Blanton {et~al.}(2017)Blanton, Bershady, Abolfathi, Albareti, Prieto,
  Almeida, Alonso-Garc{\'\i}a, Anders, Anderson, Andrews, {et~al.}}]{blanton17}
Blanton, M.~R., Bershady, M.~A., Abolfathi, B., {et~al.} 2017, The Astronomical
  Journal, 154, 28

\bibitem[{Boquien {et~al.}(2019)Boquien, Burgarella, Roehlly, Buat, Ciesla,
  Corre, Inoue, \& Salas}]{boquien19}
Boquien, M., Burgarella, D., Roehlly, Y., {et~al.} 2019, Astronomy \&
  Astrophysics, 622, A103

\bibitem[{Boyle \& Couch(1993)}]{boyle93}
Boyle, B., \& Couch, W.~J. 1993, Monthly Notices of the Royal Astronomical
  Society, 264, 604

\bibitem[{Brough {et~al.}(2013)Brough, Croom, Sharp, Hopkins, Taylor, Baldry,
  Gunawardhana, Liske, Norberg, Robotham, {et~al.}}]{brough13}
Brough, S., Croom, S., Sharp, R., {et~al.} 2013, Monthly Notices of the Royal
  Astronomical Society, 435, 2903

\bibitem[{Bruzual \& Charlot(2003)}]{bruzual03}
Bruzual, G., \& Charlot, S. 2003, Monthly Notices of the Royal Astronomical
  Society, 344, 1000

\bibitem[{Burgarella(2015)}]{burgarella15}
Burgarella, D. 2015, IAUGA, 29, 2252450

\bibitem[{Chu \& Zhu(1988)}]{chu88}
Chu, Y., \& Zhu, X. 1988, Astronomy and Astrophysics, 205, 1

\bibitem[{Cisternas {et~al.}(2010)Cisternas, Jahnke, Inskip, Kartaltepe,
  Koekemoer, Lisker, Robaina, Scodeggio, Sheth, Trump, {et~al.}}]{cisternas10}
Cisternas, M., Jahnke, K., Inskip, K.~J., {et~al.} 2010, The Astrophysical
  Journal, 726, 57

\bibitem[{Coldwell \& Lambas(2006)}]{coldwell06}
Coldwell, G.~V., \& Lambas, D.~G. 2006, Monthly Notices of the Royal
  Astronomical Society, 371, 786

\bibitem[{Colless {et~al.}(2001)Colless, Dalton, Maddox, Sutherland, Norberg,
  Cole, Bland-Hawthorn, Bridges, Cannon, Collins, {et~al.}}]{colless01}
Colless, M., Dalton, G., Maddox, S., {et~al.} 2001, Monthly Notices of the
  Royal Astronomical Society, 328, 1039

\bibitem[{Croom {et~al.}(2004)Croom, Smith, Boyle, Shanks, Miller, Outram, \&
  Loaring}]{croom04}
Croom, S.~M., Smith, R., Boyle, B., {et~al.} 2004, Monthly Notices of the Royal
  Astronomical Society, 349, 1397

\bibitem[{Da~Cunha {et~al.}(2011)Da~Cunha, Charlot, Dunne, Smith, \&
  Rowlands}]{da11}
Da~Cunha, E., Charlot, S., Dunne, L., Smith, D., \& Rowlands, K. 2011,
  Proceedings of the International Astronomical Union, 7, 292

\bibitem[{Davis {et~al.}(2018)Davis, Graham, \& Cameron}]{davis18}
Davis, B.~L., Graham, A.~W., \& Cameron, E. 2018, The Astrophysical Journal,
  869, 113

\bibitem[{Davis {et~al.}(2019)Davis, Graham, \& Cameron}]{davis19}
---. 2019, The Astrophysical Journal, 873, 85

\bibitem[{Disney {et~al.}(1995)Disney, Boyce, Blades, Boksenberg, Crane,
  Deharveng, Macchetto, Mackay, Sparks, \& Phillipps}]{disney95}
Disney, M., Boyce, P., Blades, J., {et~al.} 1995, Nature, 376, 150

\bibitem[{Draine {et~al.}(2013)Draine, Aniano, Krause, Groves, Sandstrom,
  Braun, Leroy, Klaas, Linz, Rix, {et~al.}}]{draine14}
Draine, B., Aniano, G., Krause, O., {et~al.} 2013, The Astrophysical Journal,
  780, 172

\bibitem[{Dressler(1980)}]{dressler80}
Dressler, A. 1980, The Astrophysical Journal, 236, 351

\bibitem[{Driver {et~al.}(2016)Driver, Wright, Andrews, Davies, Kafle, Lange,
  Moffett, Mannering, Robotham, Vinsen, {et~al.}}]{driver16}
Driver, S.~P., Wright, A.~H., Andrews, S.~K., {et~al.} 2016, Monthly Notices of
  the Royal Astronomical Society, 455, 3911

\bibitem[{Driver {et~al.}(2018)Driver, Andrews, Da~Cunha, Davies, Lagos,
  Robotham, Vinsen, Wright, Alpaslan, Bland-Hawthorn, {et~al.}}]{driver18}
Driver, S.~P., Andrews, S.~K., Da~Cunha, E., {et~al.} 2018, Monthly Notices of
  the Royal Astronomical Society, 475, 2891

\bibitem[{Eardley {et~al.}(2015)Eardley, Peacock, McNaught-Roberts, Heymans,
  Norberg, Alpaslan, Baldry, Bland-Hawthorn, Brough, Cluver,
  {et~al.}}]{eardley15}
Eardley, E., Peacock, J., McNaught-Roberts, T., {et~al.} 2015, Monthly Notices
  of the Royal Astronomical Society, 448, 3665

\bibitem[{Einasto {et~al.}(2005)Einasto, Tago, Einasto, Saar, Suhhonenko,
  Hein{\"a}m{\"a}ki, H{\"u}tsi, \& Tucker}]{einasto05}
Einasto, J., Tago, E., Einasto, M., {et~al.} 2005, Astronomy \& Astrophysics,
  439, 45

\bibitem[{Ellingson {et~al.}(1991)Ellingson, Yee, \& Green}]{ellingson91}
Ellingson, E., Yee, H., \& Green, R. 1991, The Astrophysical Journal, 371, 49

\bibitem[{Fisher {et~al.}(1996)Fisher, Bahcall, Kirhakos, \&
  Schneider}]{fisher96}
Fisher, K.~B., Bahcall, J.~N., Kirhakos, S., \& Schneider, D.~P. 1996, arXiv
  preprint astro-ph/9602078

\bibitem[{Fisher(1958)}]{fisher58}
Fisher, R. 1958, Edinburgh and London

\bibitem[{Gao {et~al.}(2005)Gao, Springel, \& White}]{gao05}
Gao, L., Springel, V., \& White, S.~D. 2005, Monthly Notices of the Royal
  Astronomical Society: Letters, 363, L66

\bibitem[{Gattano {et~al.}(2018)Gattano, Andrei, Coelho, Souchay, Barache, \&
  Taris}]{gattano18}
Gattano, C., Andrei, A., Coelho, B., {et~al.} 2018, Astronomy \& Astrophysics,
  614, A140

\bibitem[{Gilmour {et~al.}(2007)Gilmour, Gray, Almaini, Best, Wolf,
  Meisenheimer, Papovich, \& Bell}]{gilmour07}
Gilmour, R., Gray, M., Almaini, O., {et~al.} 2007, Monthly Notices of the Royal
  Astronomical Society, 380, 1467

\bibitem[{Gomez {et~al.}(2003)Gomez, Nichol, Miller, Balogh, Goto, Zabludoff,
  Romer, Bernardi, Sheth, Hopkins, {et~al.}}]{gomez03}
Gomez, P.~L., Nichol, R.~C., Miller, C.~J., {et~al.} 2003, The Astrophysical
  Journal, 584, 210

\bibitem[{Graham {et~al.}(2016)Graham, Ciambur, \& Soria}]{graham16}
Graham, A.~W., Ciambur, B.~C., \& Soria, R. 2016, The Astrophysical Journal,
  818, 172

\bibitem[{{Habouzit} {et~al.}(2019){Habouzit}, {Volonteri}, {Somerville},
  {Dubois}, {Peirani}, {Pichon}, \& {Devriendt}}]{habouzit2019}
{Habouzit}, M., {Volonteri}, M., {Somerville}, R.~S., {et~al.} 2019, \mnras,
  489, 1206, \dodoi{10.1093/mnras/stz2105}

\bibitem[{Hennawi {et~al.}(2006)Hennawi, Strauss, Oguri, Inada, Richards,
  Pindor, Schneider, Becker, Gregg, Hall, {et~al.}}]{hennawi06}
Hennawi, J.~F., Strauss, M.~A., Oguri, M., {et~al.} 2006, The Astronomical
  Journal, 131, 1

\bibitem[{Hintzen {et~al.}(1991)Hintzen, Romanishin, \& Valdes}]{hintzen91}
Hintzen, P., Romanishin, W., \& Valdes, F. 1991, The Astrophysical Journal,
  366, 7

\bibitem[{Hopkins {et~al.}(2013)Hopkins, Driver, Brough, Owers, Bauer,
  Gunawardhana, Cluver, Colless, Foster, Lara-L{\'o}pez, {et~al.}}]{hopkins13}
Hopkins, A.~M., Driver, S.~P., Brough, S., {et~al.} 2013, Monthly Notices of
  the Royal Astronomical Society, 430, 2047

\bibitem[{Hopkins {et~al.}(2007)Hopkins, Bundy, Hernquist, \&
  Ellis}]{hopkins07}
Hopkins, P.~F., Bundy, K., Hernquist, L., \& Ellis, R.~S. 2007, The
  Astrophysical Journal, 659, 976

\bibitem[{Hopkins {et~al.}(2006)Hopkins, Hernquist, Cox, Di~Matteo, Robertson,
  \& Springel}]{hopkins06}
Hopkins, P.~F., Hernquist, L., Cox, T.~J., {et~al.} 2006, The Astrophysical
  Journal Supplement Series, 163, 1

\bibitem[{Inoue(2011)}]{inoue11}
Inoue, A.~K. 2011, Monthly Notices of the Royal Astronomical Society, 415, 2920

\bibitem[{Karhunen {et~al.}(2014)Karhunen, Kotilainen, Falomo, \&
  Bettoni}]{karhunen14}
Karhunen, K., Kotilainen, J., Falomo, R., \& Bettoni, D. 2014, Monthly Notices
  of the Royal Astronomical Society, 441, 1802

\bibitem[{{Kayo} \& {Oguri}(2012)}]{kayo2012}
{Kayo}, I., \& {Oguri}, M. 2012, \mnras, 424, 1363,
  \dodoi{10.1111/j.1365-2966.2012.21321.x}

\bibitem[{Kocevski {et~al.}(2011)Kocevski, Faber, Mozena, Koekemoer, Nandra,
  Rangel, Laird, Brusa, Wuyts, Trump, {et~al.}}]{kocevski11}
Kocevski, D.~D., Faber, S., Mozena, M., {et~al.} 2011, The Astrophysical
  Journal, 744, 148

\bibitem[{Kormendy \& Ho(2013)}]{kormendy13}
Kormendy, J., \& Ho, L.~C. 2013, Annual Review of Astronomy and Astrophysics,
  51, 511

\bibitem[{Lewis {et~al.}(2002)Lewis, Balogh, De~Propris, Couch, Bower, Offer,
  Bland-Hawthorn, Baldry, Baugh, Bridges, {et~al.}}]{lewis02}
Lewis, I., Balogh, M., De~Propris, R., {et~al.} 2002, Monthly Notices of the
  Royal Astronomical Society, 334, 673

\bibitem[{Lietzen {et~al.}(2011)Lietzen, Hein{\"a}m{\"a}ki, Nurmi,
  Liivam{\"a}gi, Saar, Tago, Takalo, \& Einasto}]{lietzen11}
Lietzen, H., Hein{\"a}m{\"a}ki, P., Nurmi, P., {et~al.} 2011, Astronomy \&
  Astrophysics, 535, A21

\bibitem[{Lietzen {et~al.}(2009)Lietzen, Hein{\"a}m{\"a}ki, Nurmi, Tago, Saar,
  Liivam{\"a}gi, Tempel, Einasto, Einasto, Gramann, {et~al.}}]{lietzen09}
---. 2009, Astronomy \& Astrophysics, 501, 145

\bibitem[{Liske {et~al.}(2015)Liske, Baldry, Driver, Tuffs, Alpaslan, Andrae,
  Brough, Cluver, Grootes, Gunawardhana, {et~al.}}]{liske15}
Liske, J., Baldry, I.~K., Driver, S.~P., {et~al.} 2015, Monthly Notices of the
  Royal Astronomical Society, 452, 2087

\bibitem[{Loveday {et~al.}(2015)Loveday, Norberg, Baldry, Bland-Hawthorn,
  Brough, Brown, Driver, Kelvin, \& Phillipps}]{loveday15}
Loveday, J., Norberg, P., Baldry, I., {et~al.} 2015, Monthly Notices of the
  Royal Astronomical Society, 451, 1540

\bibitem[{Magorrian {et~al.}(1998)Magorrian, Tremaine, Richstone, Bender,
  Bower, Dressler, Faber, Gebhardt, Green, Grillmair, {et~al.}}]{magorrian98}
Magorrian, J., Tremaine, S., Richstone, D., {et~al.} 1998, The Astronomical
  Journal, 115, 2285

\bibitem[{Massey~Jr(1951)}]{massey51}
Massey~Jr, F.~J. 1951, Journal of the American statistical Association, 46, 68

\bibitem[{McLure \& Dunlop(2001)}]{mclure01}
McLure, R., \& Dunlop, J. 2001, Monthly Notices of the Royal Astronomical
  Society, 321, 515

\bibitem[{Miller {et~al.}(2003)Miller, Nichol, Gomez, Hopkins, \&
  Bernardi}]{miller03}
Miller, C.~J., Nichol, R.~C., Gomez, P.~L., Hopkins, A.~M., \& Bernardi, M.
  2003, The Astrophysical Journal, 597, 142

\bibitem[{Moon {et~al.}(2019)Moon, An, \& Yoon}]{moon19}
Moon, J.-S., An, S.-H., \& Yoon, S.-J. 2019, The Astrophysical Journal, 882, 14

\bibitem[{Muldrew {et~al.}(2012)Muldrew, Croton, Skibba, Pearce, Ann, Baldry,
  Brough, Choi, Conselice, Cowan, {et~al.}}]{muldrew12}
Muldrew, S.~I., Croton, D.~J., Skibba, R.~A., {et~al.} 2012, Monthly Notices of
  the Royal Astronomical Society, 419, 2670

\bibitem[{Noll {et~al.}(2009)Noll, Burgarella, Giovannoli, Buat, Marcillac, \&
  Munoz-Mateos}]{noll09}
Noll, S., Burgarella, D., Giovannoli, E., {et~al.} 2009, Astronomy \&
  Astrophysics, 507, 1793

\bibitem[{Oemler(1974)}]{oemler74}
Oemler, A. 1974, PhD thesis, California Institute of Technology

\bibitem[{Porter {et~al.}(2008)Porter, Raychaudhury, Pimbblet, \&
  Drinkwater}]{porter08}
Porter, S.~C., Raychaudhury, S., Pimbblet, K.~A., \& Drinkwater, M.~J. 2008,
  Monthly Notices of the Royal Astronomical Society, 388, 1152

\bibitem[{{Richardson} {et~al.}(2012){Richardson}, {Zheng}, {Chatterjee},
  {Nagai}, \& {Shen}}]{richardson2012}
{Richardson}, J., {Zheng}, Z., {Chatterjee}, S., {Nagai}, D., \& {Shen}, Y.
  2012, \apj, 755, 30, \dodoi{10.1088/0004-637X/755/1/30}

\bibitem[{Robotham {et~al.}(2011)Robotham, Norberg, Driver, Baldry, Bamford,
  Hopkins, Liske, Loveday, Merson, Peacock, {et~al.}}]{robotham11}
Robotham, A.~S., Norberg, P., Driver, S.~P., {et~al.} 2011, Monthly Notices of
  the Royal Astronomical Society, 416, 2640

\bibitem[{Sahu {et~al.}(2019)Sahu, Graham, \& Davis}]{sahu19}
Sahu, N., Graham, A.~W., \& Davis, B.~L. 2019, The Astrophysical Journal, 876,
  155

\bibitem[{Sanders {et~al.}(1988)Sanders, Soifer, Elias, Neugebauer, \&
  Matthews}]{sanders88}
Sanders, D., Soifer, B., Elias, J., Neugebauer, G., \& Matthews, K. 1988,
  Astrophysical Journal, 328, L35

\bibitem[{Saunders {et~al.}(2004)Saunders, Bridges, Gillingham, Haynes, Smith,
  Whittard, Churilov, Lankshear, Croom, Jones, {et~al.}}]{saunders04}
Saunders, W., Bridges, T., Gillingham, P., {et~al.} 2004, in Ground-based
  Instrumentation for Astronomy, Vol. 5492, International Society for Optics
  and Photonics, 389--400

\bibitem[{Savorgnan {et~al.}(2016)Savorgnan, Graham, Marconi, \&
  Sani}]{savorgnan16}
Savorgnan, G.~A., Graham, A.~W., Marconi, A., \& Sani, E. 2016, The
  Astrophysical Journal, 817, 21

\bibitem[{Serber {et~al.}(2006)Serber, Bahcall, M{\'e}nard, \&
  Richards}]{serber06}
Serber, W., Bahcall, N., M{\'e}nard, B., \& Richards, G. 2006, The
  Astrophysical Journal, 643, 68

\bibitem[{Shanks {et~al.}(1988)Shanks, Boyle, \& Peterson}]{shanks88}
Shanks, T., Boyle, B., \& Peterson, B. 1988, in Optical Surveys for Quasars,
  Vol.~2, 244

\bibitem[{Sharp {et~al.}(2006)Sharp, Saunders, Smith, Churilov, Correll,
  Dawson, Farrel, Frost, Haynes, Heald, {et~al.}}]{sharp06}
Sharp, R., Saunders, W., Smith, G., {et~al.} 2006, in Ground-based and Airborne
  Instrumentation for Astronomy, Vol. 6269, International Society for Optics
  and Photonics, 62690G

\bibitem[{Shattow {et~al.}(2013)Shattow, Croton, Skibba, Muldrew, Pearce, \&
  Abbas}]{shattow13}
Shattow, G.~M., Croton, D.~J., Skibba, R.~A., {et~al.} 2013, Monthly Notices of
  the Royal Astronomical Society, 433, 3314

\bibitem[{{Shen} {et~al.}(2013){Shen}, {McBride}, {White}, {Zheng}, {Myers},
  {Guo}, {Kirkpatrick}, {Padmanabhan}, {Parejko}, {Ross}, {Schlegel},
  {Schneider}, {Streblyanska}, {Swanson}, {Zehavi}, {Pan}, {Bizyaev},
  {Brewington}, {Ebelke}, {Malanushenko}, {Malanushenko}, {Oravetz}, {Simmons},
  \& {Snedden}}]{shen2013}
{Shen}, Y., {McBride}, C.~K., {White}, M., {et~al.} 2013, \apj, 778, 98,
  \dodoi{10.1088/0004-637X/778/2/98}

\bibitem[{Skibba {et~al.}(2009)Skibba, Bamford, Nichol, Lintott, Andreescu,
  Edmondson, Murray, Raddick, Schawinski, Slosar, {et~al.}}]{skibba09}
Skibba, R.~A., Bamford, S.~P., Nichol, R.~C., {et~al.} 2009, Monthly Notices of
  the Royal Astronomical Society, 399, 966

\bibitem[{Smith {et~al.}(2004)Smith, Saunders, Bridges, Churilov, Lankshear,
  Dawson, Correll, Waller, Haynes, \& Frost}]{smith04}
Smith, G.~A., Saunders, W., Bridges, T., {et~al.} 2004, in Ground-based
  Instrumentation for Astronomy, Vol. 5492, International Society for Optics
  and Photonics, 410--420

\bibitem[{S{\"o}chting {et~al.}(2002)S{\"o}chting, Clowes, \&
  Campusano}]{sochting02}
S{\"o}chting, I.~K., Clowes, R.~G., \& Campusano, L.~E. 2002, Monthly notices
  of the royal astronomical society, 331, 569

\bibitem[{Stockton(1978)}]{stockton78}
Stockton, A. 1978, The Astrophysical Journal, 223, 747

\bibitem[{Stott {et~al.}(2020)Stott, Bielby, Cullen, Burchett, Tejos,
  Fumagalli, Crain, Morris, Amos, Bower, {et~al.}}]{stott20}
Stott, J., Bielby, R., Cullen, F., {et~al.} 2020, Monthly Notices of the Royal
  Astronomical Society, 497, 3083

\bibitem[{Taylor {et~al.}(2011)Taylor, Hopkins, Baldry, Brown, Driver, Kelvin,
  Hill, Robotham, Bland-Hawthorn, Jones, {et~al.}}]{taylor11}
Taylor, E.~N., Hopkins, A.~M., Baldry, I.~K., {et~al.} 2011, Monthly Notices of
  the Royal Astronomical Society, 418, 1587

\bibitem[{Treister {et~al.}(2012)Treister, Schawinski, Urry, \&
  Simmons}]{treister12}
Treister, E., Schawinski, K., Urry, C., \& Simmons, B.~D. 2012, The
  Astrophysical Journal Letters, 758, L39

\bibitem[{Veilleux {et~al.}(2002)Veilleux, Kim, \& Sanders}]{veilleux02}
Veilleux, S., Kim, D.-C., \& Sanders, D. 2002, The Astrophysical Journal
  Supplement Series, 143, 315

\bibitem[{{Villforth} {et~al.}(2012){Villforth}, {Sarajedini}, \&
  {Koekemoer}}]{villforth12}
{Villforth}, C., {Sarajedini}, V., \& {Koekemoer}, A. 2012, \mnras, 426, 360,
  \dodoi{10.1111/j.1365-2966.2012.21732.x}

\bibitem[{Villforth {et~al.}(2014)Villforth, Hamann, Rosario, Santini, McGrath,
  Wel, Chang, Guo, Dahlen, Bell, {et~al.}}]{villforth14}
Villforth, C., Hamann, F., Rosario, D., {et~al.} 2014, Monthly Notices of the
  Royal Astronomical Society, 439, 3342

\bibitem[{Von Der~Linden {et~al.}(2007)Von Der~Linden, Best, Kauffmann, \&
  White}]{von07}
Von Der~Linden, A., Best, P.~N., Kauffmann, G., \& White, S.~D. 2007, Monthly
  Notices of the Royal Astronomical Society, 379, 867

\bibitem[{Wang {et~al.}(2011)Wang, Mo, Jing, Yang, \& Wang}]{wang11}
Wang, H., Mo, H., Jing, Y., Yang, X., \& Wang, Y. 2011, Monthly Notices of the
  Royal Astronomical Society, 413, 1973

\bibitem[{{Wang} {et~al.}(2015){Wang}, {Viero}, {Ross}, {Asboth},
  {B{\'e}thermin}, {Bock}, {Clements}, {Conley}, {Cooray}, {Farrah}, {Hajian},
  {Han}, {Lagache}, {Marsden}, {Myers}, {Norberg}, {Oliver}, {Page},
  {Symeonidis}, {Schulz}, {Wang}, \& {Zemcov}}]{wang2014}
{Wang}, L., {Viero}, M., {Ross}, N.~P., {et~al.} 2015, \mnras, 449, 4476,
  \dodoi{10.1093/mnras/stv559}

\bibitem[{Wijesinghe {et~al.}(2012)Wijesinghe, Hopkins, Brough, Taylor,
  Norberg, Bauer, Brown, Cameron, Conselice, Croom, {et~al.}}]{wijesinghe12}
Wijesinghe, D., Hopkins, A.~M., Brough, S., {et~al.} 2012, Monthly Notices of
  the Royal Astronomical Society, 423, 3679

\bibitem[{Wright {et~al.}(2017)Wright, Robotham, Driver, Alpaslan, Andrews,
  Baldry, Bland-Hawthorn, Brough, Brown, Colless, {et~al.}}]{wright17}
Wright, A., Robotham, A., Driver, S., {et~al.} 2017, Monthly Notices of the
  Royal Astronomical Society, 470, 283

\bibitem[{Yee \& Green(1984)}]{yee84}
Yee, H., \& Green, R.~F. 1984, The Astrophysical Journal, 280, 79

\bibitem[{York {et~al.}(2000)York, Adelman, Anderson~Jr, Anderson, Annis,
  Bahcall, Bakken, Barkhouser, Bastian, Berman, {et~al.}}]{york00}
York, D.~G., Adelman, J., Anderson~Jr, J.~E., {et~al.} 2000, The Astronomical
  Journal, 120, 1579

\bibitem[{Zhang {et~al.}(2013)Zhang, Wang, Wang, \& Zhou}]{zhang13}
Zhang, S., Wang, T., Wang, H., \& Zhou, H. 2013, The Astrophysical Journal,
  773, 175

\end{thebibliography}
\bibliographystyle{aasjournal}



\appendix

\title{List of the 205 GAMA quasar targets used in the study.}

\begin{center}
\begin{longtable}{l|l|l|l}
\caption{List of the 205 GAMA quasar targets used in the study.} \label{tab:data_full} \\

\hline \multicolumn{1}{c|}{CATAID} & \multicolumn{1}{c|}{RA (deg)} & \multicolumn{1}{c|}{Dec (deg)} & \multicolumn{1}{c}{$z$} \\ \hline 
\endfirsthead

\multicolumn{4}{c}%
{{\textbf{\tablename\ \thetable{}} -- continued from previous page}} \\

\hline \multicolumn{1}{c|}{CATAID} & \multicolumn{1}{c|}{RA (deg)} & \multicolumn{1}{c|}{Dec (deg)} & \multicolumn{1}{c}{$z$} \\ \hline 
\endhead

\hline \multicolumn{4}{r}{{\textit{Continued on next page}}} \\ 
\endfoot

\hline 
\endlastfoot
        729886 & 130.2953 & 2.4978 & 0.3320 \\
        386308 & 131.3965 & 2.1527 & 0.1971 \\
        599408 & 131.3974 & 0.2721 & 0.2611 \\
        202846 & 131.4814 & -0.2153 & 0.2750 \\
        209050 & 131.6569 & 0.0068 & 0.2575 \\
        549519 & 131.6925 & -0.5092 & 0.2548 \\
        743886 & 131.9999 & -0.5593 & 0.2678 \\
        323581 & 132.2357 & 1.6133 & 0.3499 \\
        381362 & 132.4181 & 1.7773 & 0.3289 \\
        730234 & 133.1228 & 2.7872 & 0.3345 \\
        371891 & 133.3412 & 1.0380 & 0.2140 \\
        549878 & 133.3905 & -0.5153 & 0.1883 \\
        744010 & 133.6926 & -0.5638 & 0.2678 \\
        323919 & 133.7043 & 1.6836 & 0.2614 \\
        727391 & 133.9371 & 2.2504 & 0.2607 \\
        376679 & 134.6499 & 1.5304 & 0.1068 \\
        346711 & 134.7665 & 2.0844 & 0.1892 \\
        719471 & 134.8152 & 1.3001 & 0.2821 \\
        279031 & 135.0242 & 0.9765 & 0.2504 \\
        346835 & 135.1853 & 2.0447 & 0.1287 \\
        376800 & 135.2267 & 1.4348 & 0.2607 \\
        600408 & 135.4954 & 0.3872 & 0.1963 \\
        215842 & 135.5890 & 0.5725 & 0.3262 \\
        215843 & 135.6269 & 0.6044 & 0.1982 \\
        347016 & 135.7827 & 2.1978 & 0.3291 \\
        372457 & 135.7938 & 1.1880 & 0.2049 \\
        372453 & 135.8129 & 1.1064 & 0.1217 \\
        372542 & 136.2413 & 1.0731 & 0.2789 \\
        719292 & 136.3016 & 1.3006 & 0.2489 \\
        377086 & 136.4325 & 1.5158 & 0.1018 \\
        210075 & 136.5215 & 0.0479 & 0.2002 \\
        518519 & 136.6816 & 2.7868 & 0.2539 \\
        377246 & 136.9312 & 1.5578 & 0.1643 \\
        550701 & 136.9690 & -0.5175 & 0.2000 \\
        600710 & 137.1114 & 0.3604 & 0.1626 \\
        210251 & 137.3017 & 0.0394 & 0.2936 \\
        709902 & 137.3404 & 0.6021 & 0.2905 \\
        347434 & 137.7302 & 2.2826 & 0.1828 \\
        387857 & 137.9571 & 2.4883 & 0.1546 \\
        623320 & 138.3153 & 0.7457 & 0.2253 \\
        745714 & 138.5242 & -0.0026 & 0.3007 \\
        302636 & 138.6557 & 1.4398 & 0.1870 \\
        302641 & 138.7460 & 1.4421 & 0.1977 \\
        365307 & 138.7482 & 2.5864 & 0.2332 \\
        747874 & 138.8033 & 0.3069 & 0.3077 \\
        721180 & 138.8082 & 1.7320 & 0.2023 \\
        377827 & 139.0578 & 1.7095 & 0.2444 \\
        382919 & 139.1085 & 2.0231 & 0.1352 \\
        743963 & 139.1756 & -0.4406 & 0.3154 \\
        708278 & 139.2059 & 0.0088 & 0.2225 \\
        388235 & 139.2300 & 2.5121 & 0.1391 \\
        302859 & 139.6826 & 1.5157 & 0.2034 \\
        302896 & 139.7798 & 1.3462 & 0.2870 \\
        551411 & 140.2441 & -0.5068 & 0.3281 \\
        383263 & 140.7556 & 2.1914 & 0.3424 \\
        183958 & 174.3836 & -1.5873 & 0.3415 \\
        402343 & 174.6727 & 1.9270 & 0.2571 \\
        53744 & 174.7903 & -0.2690 & 0.1348 \\
        534762 & 174.9536 & -0.9086 & 0.1901 \\
        30401 & 175.0181 & -1.0910 & 0.3472 \\
        690012 & 175.0295 & -1.2052 & 0.2279 \\
        402515 & 175.6541 & 1.9315 & 0.1200 \\
        30518 & 175.7006 & -1.1823 & 0.2235 \\
        559141 & 175.8973 & -0.4951 & 0.1715 \\
        136689 & 175.9249 & -1.7429 & 0.1051 \\
        559310 & 176.4752 & -0.6120 & 0.2810 \\
        271882 & 176.7834 & 1.3270 & 0.2298 \\
        718373 & 177.0644 & 1.6445 & 0.2686 \\
        718466 & 177.2014 & 1.6762 & 0.1598 \\
        30887 & 177.2069 & -1.1696 & 0.1058 \\
        583923 & 177.3678 & -0.0785 & 0.1759 \\
        69923 & 177.5983 & 0.1442 & 0.1271 \\
        7454 & 177.7092 & 0.7516 & 0.1394 \\
        7512 & 177.9093 & 0.8296 & 0.1952 \\
        143873 & 177.9862 & -1.3002 & 0.1706 \\
        746435 & 178.1458 & -0.0952 & 0.1287 \\
        39798 & 178.3365 & -0.7605 & 0.1808 \\
        7618 & 178.6182 & 0.6877 & 0.2286 \\
        713630 & 178.8723 & 1.4926 & 0.2404 \\
        22727 & 178.8896 & 1.1251 & 0.1970 \\
        272384 & 179.1655 & 1.4334 & 0.2275 \\
        70269 & 179.2836 & 0.0471 & 0.1314 \\
        748358 & 179.4811 & 0.2844 & 0.2607 \\
        692601 & 179.4947 & -0.3725 & 0.2599 \\
        7847 & 179.6604 & 0.7087 & 0.1998 \\
        31502 & 179.9447 & -1.0807 & 0.1321 \\
        691054 & 180.0587 & -0.7774 & 0.1794 \\
        718498 & 180.5202 & 1.8356 & 0.1390 \\
        185507 & 180.6115 & -1.4876 & 0.1504 \\
        230680 & 180.8006 & 1.8893 & 0.2960 \\
        99085 & 180.9243 & 1.0019 & 0.2417 \\
        185590 & 180.9561 & -1.5817 & 0.2833 \\
        560666 & 181.0609 & -0.5026 & 0.1686 \\
        700164 & 181.1375 & -1.7079 & 0.2023 \\
        178669 & 181.2447 & -1.9659 & 0.2880 \\
        220373 & 181.3178 & 1.5856 & 0.1119 \\
        713763 & 181.7788 & 1.5835 & 0.2107 \\
        767417 & 181.8868 & 0.5586 & 0.2779 \\
        55376 & 181.8961 & -0.2640 & 0.1104 \\
        714379 & 182.5765 & 1.9017 & 0.2155 \\
        692590 & 182.6100 & -0.3604 & 0.2915 \\
        561045 & 182.6407 & -0.4745 & 0.2862 \\
        40708 & 182.6815 & -0.6520 & 0.3311 \\
        585421 & 182.8242 & -0.0368 & 0.1811 \\
        751044 & 182.9274 & 1.0603 & 0.2933 \\
        561186 & 183.1355 & -0.4691 & 0.2750 \\
        231306 & 183.1516 & 1.9675 & 0.1894 \\
        696654 & 183.2100 & 0.8602 & 0.3075 \\
        186167 & 183.3883 & -1.5391 & 0.1977 \\
        561225 & 183.4205 & -0.5802 & 0.2534 \\
        742715 & 183.4323 & -1.0007 & 0.3279 \\
        85698 & 183.6570 & 0.4764 & 0.1849 \\
        749954 & 183.6584 & 0.7420 & 0.2497 \\
        748263 & 183.7275 & 0.2488 & 0.3145 \\
        55958 & 184.1544 & -0.3778 & 0.1573 \\
        703395 & 184.2737 & -1.9729 & 0.2936 \\
        537252 & 184.6102 & -0.8670 & 0.1531 \\
        700865 & 184.6898 & -1.4294 & 0.2384 \\
        138906 & 184.7864 & -1.8081 & 0.1030 \\
        749971 & 184.8075 & 0.6809 & 0.2857 \\
        704000 & 185.0398 & -1.5336 & 0.2877 \\
        56248 & 185.1374 & -0.2428 & 0.2119 \\
        71576 & 185.2808 & 0.0475 & 0.1572 \\
        290243 & 185.3397 & 1.6716 & 0.2671 \\
        586591 & 185.5744 & -0.1288 & 0.1729 \\
        692981 & 211.6254 & -0.3277 & 0.1066 \\
        91404 & 211.8465 & 0.4756 & 0.1054 \\
        567506 & 211.9747 & -0.5052 & 0.1672 \\
        318702 & 212.0668 & 1.9245 & 0.1661 \\
        463260 & 212.2304 & -1.2551 & 0.1951 \\
        617634 & 212.5153 & 0.2140 & 0.1358 \\
        736161 & 212.5297 & -1.3341 & 0.3200 \\
        318826 & 212.7065 & 1.8597 & 0.2011 \\
        567768 & 213.1445 & -0.5834 & 0.1266 \\
        463453 & 213.1615 & -1.3427 & 0.1396 \\
        278136 & 213.1826 & 1.1406 & 0.2266 \\
        463493 & 213.3114 & -1.2615 & 0.1208 \\
        62528 & 213.3634 & -0.2431 & 0.1254 \\
        62700 & 213.6248 & -0.3862 & 0.3225 \\
        567908 & 213.6294 & -0.5119 & 0.1378 \\
        238270 & 213.7273 & 1.5663 & 0.2699 \\
        568049 & 213.8313 & -0.5060 & 0.1345 \\
        463622 & 213.8572 & -1.2313 & 0.1480 \\
        717517 & 213.8672 & 1.1114 & 0.2655 \\
        227569 & 214.1663 & 1.2748 & 0.1383 \\
        296974 & 214.1922 & 1.5203 & 0.1164 \\
        238492 & 214.4532 & 1.6362 & 0.2523 \\
        544256 & 214.5847 & -0.9983 & 0.2538 \\
        511675 & 215.4922 & -1.1324 & 0.1913 \\
        106412 & 216.0103 & 1.0248 & 0.2952 \\
        744812 & 216.0158 & -0.4494 & 0.1510 \\
        48183 & 216.1419 & -0.6331 & 0.1583 \\
        297516 & 216.2062 & 1.3982 & 0.1076 \\
        297575 & 216.2880 & 1.3198 & 0.2000 \\
        748815 & 216.4413 & 0.3785 & 0.3257 \\
        697359 & 216.7032 & 0.8898 & 0.2195 \\
        492449 & 216.7125 & -1.2981 & 0.2612 \\
        63765 & 216.8610 & -0.3306 & 0.2219 \\
        593523 & 216.9312 & -0.0705 & 0.3065 \\
        593582 & 217.0008 & -0.1741 & 0.3330 \\
        485691 & 217.0360 & -1.7521 & 0.3245 \\
        492554 & 217.1141 & -1.3127 & 0.2813 \\
        239321 & 217.3239 & 1.6207 & 0.3112 \\
        228508 & 217.5981 & 1.1361 & 0.1141 \\
        544920 & 217.6153 & -0.9375 & 0.3179 \\
        593722 & 217.6259 & -0.1875 & 0.1031 \\
        691892 & 217.7569 & -0.7144 & 0.3337 \\
        298051 & 218.1552 & 1.3475 & 0.1382 \\
        239599 & 218.4509 & 1.4938 & 0.2111 \\
        48802 & 218.6032 & -0.6489 & 0.2826 \\
        239709 & 218.6586 & 1.6463 & 0.1350 \\
        48873 & 218.7956 & -0.6394 & 0.2905 \\
        239790 & 218.9835 & 1.6269 & 0.2770 \\
        569415 & 219.1034 & -0.4848 & 0.3248 \\
        505311 & 219.1830 & -1.9370 & 0.1290 \\
        512558 & 219.2550 & -1.0717 & 0.2852 \\
        694503 & 219.2672 & 0.1181 & 0.1402 \\
        745122 & 219.3601 & -0.4410 & 0.2600 \\
        64474 & 219.3659 & -0.3954 & 0.1379 \\
        298359 & 219.3827 & 1.3161 & 0.3423 \\
        505383 & 219.6715 & -1.9551 & 0.2970 \\
        594304 & 219.6981 & -0.1348 & 0.1040 \\
        16696 & 219.7248 & 0.6639 & 0.3390 \\
        619445 & 219.8676 & 0.2606 & 0.3390 \\
        493251 & 220.1304 & -1.2969 & 0.1382 \\
        251811 & 220.5073 & 1.9511 & 0.1399 \\
        321011 & 220.6326 & 1.7315 & 0.2801 \\
        107532 & 220.6982 & 0.9256 & 0.2428 \\
        512888 & 220.8322 & -1.0660 & 0.2915 \\
        695942 & 220.9244 & 0.4229 & 0.2624 \\
        240161 & 220.9990 & 1.4990 & 0.2956 \\
        107651 & 221.5804 & 0.8884 & 0.1372 \\
        743291 & 221.6097 & -0.8789 & 0.3457 \\
        513114 & 221.7332 & -1.1243 & 0.2052 \\
        107713 & 221.9459 & 0.9217 & 0.2964 \\
        619879 & 221.9670 & 0.4067 & 0.2018 \\
        695970 & 221.9869 & 0.5560 & 0.2656 \\
        745053 & 222.0521 & -0.5110 & 0.3192 \\
        65202 & 222.0603 & -0.2954 & 0.2883 \\
        691908 & 222.3770 & -0.7962 & 0.2525 \\
        513244 & 222.5800 & -1.1132 & 0.1195 \\
        594989 & 222.8459 & -0.1072 & 0.1386 \\
        252245 & 222.8947 & 1.7865 & 0.2674 \\
        107854 & 223.0071 & 0.8445 & 0.3156 \\
        620120 & 223.2870 & 0.3173 & 0.3138 \\
\end{longtable}
\end{center}

\end{document}